\documentclass[prc,reprint,superscriptaddress]{revtex4-2} 

\newcommand{\mc}[1]{\multicolumn{1}{c}{#1}}
\newcommand{\nuc}[2]{$^{#1}$#2}
\newcommand{\rme}[3]{\langle{#1} \lVert{#2} \rVert{#3} \rangle}

\usepackage{amsmath}
\usepackage{booktabs}
\usepackage{dcolumn}
\usepackage{diagbox}
\usepackage[pdftex]{graphics}
\usepackage{hyperref}
\usepackage{microtype}
\usepackage{pgfplots}
\usepgfplotslibrary{groupplots}
\pgfplotsset{compat=1.8}
\usepackage{tikz}
\usetikzlibrary{calc}
\usetikzlibrary{external}
\usetikzlibrary{math}
\usetikzlibrary{matrix}

\pgfplotsset{colormap={inferno}{
			rgb (0) = (0.001462, 0.000466, 0.013866),
			rgb (15) = (0.037668, 0.025921, 0.132232),
			rgb (30) = (0.116656, 0.047574, 0.272321),
			rgb (45) = (0.217949, 0.036615, 0.383522),
			rgb (60) = (0.316282, 0.053490, 0.425116),
			rgb (75) = (0.410113, 0.087896, 0.433098),
			rgb (90) = (0.503493, 0.121575, 0.423356),
			rgb (105) = (0.596940, 0.154848, 0.398125),
			rgb (120) = (0.688653, 0.192239, 0.357603),
			rgb (135) = (0.775059, 0.239667, 0.303526),
			rgb (150) = (0.851384, 0.302260, 0.239636),
			rgb (165) = (0.912966, 0.381636, 0.169755),
			rgb (180) = (0.956852, 0.475356, 0.094695),
			rgb (195) = (0.981895, 0.579392, 0.026250),
			rgb (210) = (0.987464, 0.690366, 0.079990),
			rgb (225) = (0.973088, 0.805409, 0.216877),
			rgb (240) = (0.947594, 0.917399, 0.410665),
			rgb (255) = (0.988362, 0.998364, 0.644924),
}}

\hypersetup{
	colorlinks   = true,
	citecolor    = blue,
	urlcolor     = blue,
	linkcolor   = blue,
}

\begin{document}

\title{Resolving anomalous collectivity in the \texorpdfstring{\boldmath{$4_1^+ \to 2_1^+$}}{4₁⁺ → 2₁⁺} transition of \texorpdfstring{\nuc{\bf{58}}{Fe}}{⁵⁸Fe}}

\author{J.~A.~Woodside}
\affiliation{Department of Nuclear Physics and Accelerator Applications, Research School of Physics, The Australian National University, Canberra, ACT 2601, Australia}

\author{B.~J.~Coombes}
\affiliation{Department of Nuclear Physics and Accelerator Applications, Research School of Physics, The Australian National University, Canberra, ACT 2601, Australia}

\author{A.~E.~Stuchbery}
\affiliation{Department of Nuclear Physics and Accelerator Applications, Research School of Physics, The Australian National University, Canberra, ACT 2601, Australia}

\author{A.~J.~Mitchell}
\affiliation{Department of Nuclear Physics and Accelerator Applications, Research School of Physics, The Australian National University, Canberra, ACT 2601, Australia}

\author{M.~Reece}
\affiliation{Department of Nuclear Physics and Accelerator Applications, Research School of Physics, The Australian National University, Canberra, ACT 2601, Australia}

\author{G.~J.~Lane}
\affiliation{Department of Nuclear Physics and Accelerator Applications, Research School of Physics, The Australian National University, Canberra, ACT 2601, Australia}

\author{T.~J.~Gray}
\affiliation{Physics Division, Oak Ridge National Laboratory, Oak Ridge, TN 37831, USA}
\affiliation{Department of Physics and Astronomy, University of Tennessee, Knoxville, TN 37966, USA}

\author{G. Pasqualato}
\affiliation{CEA, LIST, Laboratoire National Henri Becquerel (LNE-LNHB), F-91120 Palaiseau, France}
\affiliation{Universit\'e Paris-Saclay, CNRS/IN2P3, IJCLab, 91405 Orsay, France}

\author{L.~J.~McKie}
\affiliation{Department of Nuclear Physics and Accelerator Applications, Research School of Physics, The Australian National University, Canberra, ACT 2601, Australia}

\author{N.~J.~Spinks}
\affiliation{Department of Nuclear Physics and Accelerator Applications, Research School of Physics, The Australian National University, Canberra, ACT 2601, Australia}

\date{\today}

\begin{abstract}
	The low-excitation states of atomic nuclei in the region around the $N = Z = 28$ shell closure are generally well described by the shell model.
	Most experimental observables in the iron isotopes \nuc{56}{Fe}, \nuc{58}{Fe}, and \nuc{60}{Fe} ($Z = 26$; $N=30$, $32$, $34$) support a shell-model description.
	However, the lifetimes of the $4_1^+$ state in \nuc{58}{Fe} in the literature result in a reduced transition strength that deviates markedly from shell-model predictions.
	There are three independent measurements, all in agreement and all based on the Doppler Shift Attenuation Method (DSAM) or Doppler-Broadened Line Shape method (DBLS).
	In this work, Coulomb-excitation measurements were performed on \nuc{56}{Fe} and \nuc{58}{Fe} beams to determine the ratios $B{\left(E2; 4_1^+ \to 2_1^+\right)}/B{\left(E2; 2_1^+ \to 0_1^+\right)}$.
	Thus, $B{\left(E2; 4_1^+ \to 2_1^+\right)}$ is determined relative to the known $B{\left(E2; 2_1^+ \to 0_1^+\right)}$ values.
	For \nuc{56}{Fe}, $B{\left(E2; 4_1^+ \to 2_1^+\right)} = 23(4)$~W.u., agreeing with the adopted value.
	However, for \nuc{58}{Fe}, the $B{\left(E2; 4_1^+ \to 2_1^+\right)}$ values obtained (for the various combinations of matrix element signs that could not be firmly established) are all significantly lower than the value derived from the previous lifetime measurements, and are in accord with shell-model calculations.
	The 1978 DSAM measurement of Bolotin \textit{et al.} [\href{https://doi.org/10.1016/0375-9474(78)90503-1}{Nucl.\ Phys.\ A 311, 75 (1978)}] has been re-examined.
	The discrepancy between that measurement and the Coulomb-excitation measurement can be ascribed to the Lindhard-Scharff-Schi{\o}tt (LSS) electronic stopping powers adopted for the DSAM analysis, which considerably overestimate contemporary values.
	Evidently, lifetime measurements from that era that are based on LSS stopping powers should be used with caution.
	The revised lifetime data, incorporating current stopping powers, are compared with shell-model calculations.
\end{abstract}

\maketitle

\section{Introduction\label{sec:introduction}}
\begin{figure*}[!ht]
	\includegraphics[width=17.2cm]{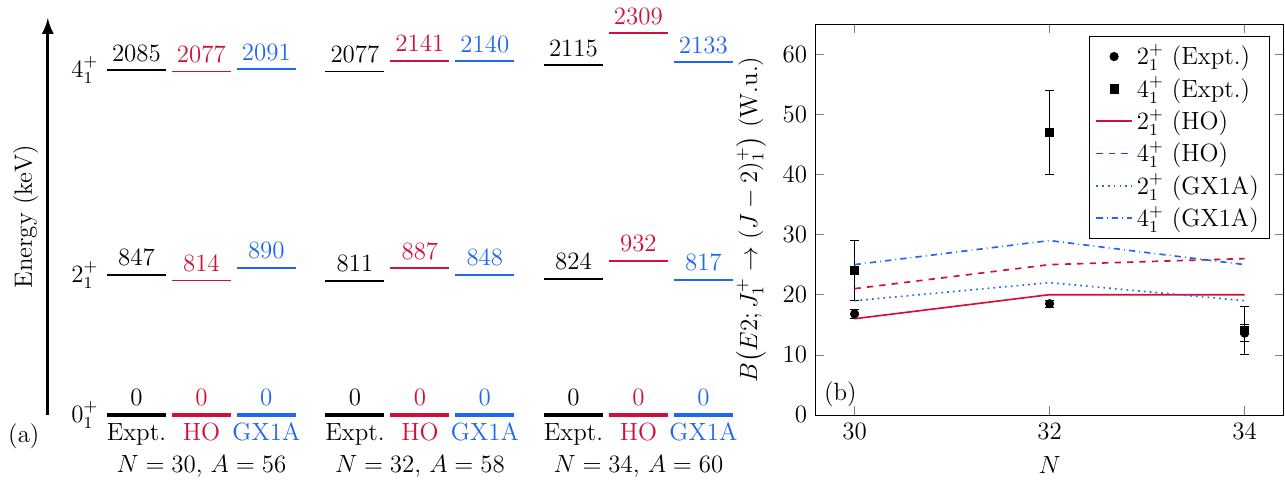}
	\caption{
		Experimental and calculated yrast-band (a) energies and (b) reduced $E2$ transition strengths in even-even iron isotopes above the $N = 28$ closed shell.
		Data are taken from Refs.~\cite{Junde2011,Nesaraja2010,Browne2013}.
		Shell-model calculations are shown for the HO~\cite{Horie1971,Horie1973} basis space and interactions, which have a $^{48}$Ca core, and for the larger $fp$ basis, which has a $^{40}$Ca core, and uses the GXPF1A~\cite{Honma2004,Honma2005} interaction (denoted GX1A in the figure).
		The recommended core-polarization charges are used, $\delta e = 1.0$ for the HO calculations and $\delta e = 0.5$ for the GXPF1A calculations (thus the effective charges for protons and neutrons are $e_{\pi}=1+\delta e$ and $e_{\nu}=\delta e$.)
		The experimental $B{\left(E2; 4_1^+ \to 2_1^+\right)}$ value for \nuc{58}{Fe} ($N = 32$) from the literature~\cite{Nesaraja2010} is significantly enhanced in comparison to neighboring isotopes, and in strong disagreement with both shell-model calculations.\label{fig:systematics}
	}
\end{figure*}

Atomic nuclei in the $N = Z = 28$ region have long been considered to be well described by the shell model, with generally good agreement between measured properties and shell-model predictions.
Iron ($Z = 26$) lies two protons below the $Z = 28$ shell closure.
The focus here is on the isotopes above $N = 28$: $^{56}$Fe$_{30}$, $^{58}$Fe$_{32}$, and $^{60}$Fe$_{34}$.
As a first approximation, the protons occupy the $0f_{7/2}$ shell while the neutrons occupy the $1p_{3/2}$, $0f_{5/2}$, and $1p_{1/2}$ orbits.
This basis space and an associated set of two-body interactions are denoted `HO' in the \textsc{NuShellX} distribution~\cite{Brown2014}.
Horie and Ogawa obtained a reasonably good description of the level schemes and electromagnetic properties of these nuclei and their neighbors with this valence space~\cite{Horie1971,Horie1973}.
More recent large-basis shell-model calculations consider the full $fp$ shell, i.e., a \nuc{40}{Ca} core with both protons and neutrons in the $0f_{7/2}$, $1p_{3/2}$, $0f_{5/2}$, and $1p_{1/2}$ orbits.
A number of interactions have been developed for this basis space; the GXPF1A interaction~\cite{Honma2004,Honma2005} gives a better description near $^{56}$Ni ($Z=N=28$) than its predecessors KB3~\cite{Poves1981} and KB3G~\cite{Poves2001}.
Fig.~\ref{fig:systematics}a compares the experimental energies for the $2_1^+$ and $4_1^+$ states in \nuc{56,58,60}{Fe} with shell-model calculations.
Both the limited-basis Horie and Ogawa calculations and calculations in the full $fp$ basis with the GXPF1A interactions (denoted GX1A in the \textsc{NuShellX} distribution) agree well with the experimental energies.

The comparison of $B(E2)$ reduced transition strengths in Fig.~\ref{fig:systematics}b also shows good agreement between theory and experiment, with the exception of the experimental $B{\left(E2; 4^+_1 \to 2^+_1\right)}$ value in \nuc{58}{Fe}, which is considerably stronger than expected.
The development of collective features in \nuc{56}{Fe} and \nuc{58}{Fe} has been suggested and investigated.
For example, the possibility of an abrupt change in shape to rotational structures in \nuc{58}{Fe} was evaluated in the HO basis space by McGrory and Raman~\cite{McGrory1979}.
Much later, Kotila and Lenzi investigated emerging collectivity in the Cr and Fe isotopes towards $N=40$~\cite{Kotila2014}, based on shell-model and interacting-boson-model approaches.

However, despite suggestions that collective structure may be emerging in these isotopes, an enhanced $B{\left(E2; 4^+_1 \to 2^+_1\right)}$ in \nuc{58}{Fe} is not expected.
The experimental ratio~\cite{Nesaraja2010} ${B{\left(E2; 4_1^+ \to 2_1^+\right)} / B{\left(E2; 2_1^+ \to 0_1^+\right)} = 2.5(4)}$ is near the vibrational model prediction of $2$, whereas the ratios for the $2^+_2 \to 2^+_1$ and $0^+_2 \to 2^+_1$ transitions of $0.5(2)$ and $0.146(5)$, respectively, are well below the vibrational estimate~\cite{Bohr1975}.

The study of Klintefjord \textit{et al.}~\cite{Klintefjord2017}, in which lifetimes were measured in \nuc{62,64}{Fe} and comparison made with large-basis shell-model calculations, also shows that the \nuc{58}{Fe} $B{\left(E2; 4_1^+ \to 2_1^+\right)}$ is an exception.
It is out of step with experimental systematics and in strong disagreement with the shell-model calculations, which otherwise describe the data well (see Fig.~7 of Ref.~\cite{Klintefjord2017}).

These observations call into question the experimental value of $B{\left(E2; 4_1^+ \to 2_1^+\right)}$ in \nuc{58}{Fe}.
The current adopted value is derived from Doppler Shift Attenuation Method (DSAM) and Doppler Broadened Line Shape (DBLS) lifetime measurements, of which there are three~\cite{Cavallaro1977,Bolotin1978,Kosyak1983}.
The previous measurements were performed independently, and the $4_1^+$ state was populated using two different reactions.
The resulting $B(E2)$ values all deviate from shell-model predictions, as does the adopted lifetime in the evaluated data~\cite{Nesaraja2010}, which is the weighted average of the three.
In view of the considerable discrepancy between these measured $4_1^+$ lifetimes and theoretical expectations, a new measurement of $B{\left(E2; 4_1^+ \to 2_1^+\right)}$ in $^{58}$Fe that avoids Doppler-shift lifetime methods is called for.
To this end, a measurement of the $B(E2)$ via multistep Coulomb-excitation was performed.

An experimental setup for performing multistep Coulomb-excitation measurements with particle-$\gamma$ coincidence spectroscopy has recently been established at the Australian Heavy Ion Accelerator Facility~\cite{Reece2025}.
The present results represent the first beam-excitation measurements performed using the new experimental setup.
In order to verify the measurement and analysis procedures a Coulomb-excitation measurement on \nuc{56}{Fe}, where the relevant $B(E2)$ values are in agreement with the shell model, was also performed under the same conditions.


A separate Coulomb-excitation study of \nuc{58}{Fe} aimed at obtaining a more comprehensive set of $E2$ matrix elements, including a precise measurement of $Q(2_1^+)$, has been performed at Orsay and is under analysis~\cite{PasqualatoUnpublished}.
The present paper is focused on the anomalous $B{\left(E2; 4_1^+ \to 2_1^+\right)}$ value and on a re-examination of the DSAM lifetime data.

The paper is arranged as follows:
Experimental details are given in Section~\ref{sec:experiment}, followed by the Coulomb-excitation analysis in Section~\ref{sec:analysis-coulex}.
Section~\ref{sec:analysis-dsam} re-examines the DSAM measurement of Bolotin \textit{et al.} and includes a full re-evaluation of the lifetimes obtained in that measurement based on contemporary stopping powers.
The discussion in Section~\ref{sec:discussion} begins with the experimental problem concerning the reliability of dated DSAM and DBLS measurements.
The discussion then continues with a comparison to the shell model using the reduced $E2$ transition strengths based on the revised lifetimes.
The concluding remarks include a short summary, and emphasize the need for caution about the accuracy of lifetime measurements that may be based on unreliable stopping powers.

\section{Experimental Details\label{sec:experiment}}
The measurements were performed at the Heavy Ion Accelerator Facility at the Australian National University~\cite{Ophel1974,Stuchbery2020}.
Beams of 220-MeV \nuc{56}{Fe} and \nuc{58}{Fe} were delivered by the 14UD Pelletron accelerator with intensities of $\approx 1$~pnA.
These beams were incident on a $718$-$\mu$g/cm$^2$-thick target of \nuc{197}{Au}, placed with its normal direction at a $20^\circ$ angle to the beam axis.
The target thickness was determined by area-weight measurements and Rutherford backscattering spectrometry.
The `safe' energies for Coulomb excitation~\cite{Cline1969} of these beam-target combinations are 222.7~MeV for \nuc{56}{Fe} and 223.7~MeV for \nuc{58}{Fe}.

\begin{figure}
	\includegraphics[width=8.6cm]{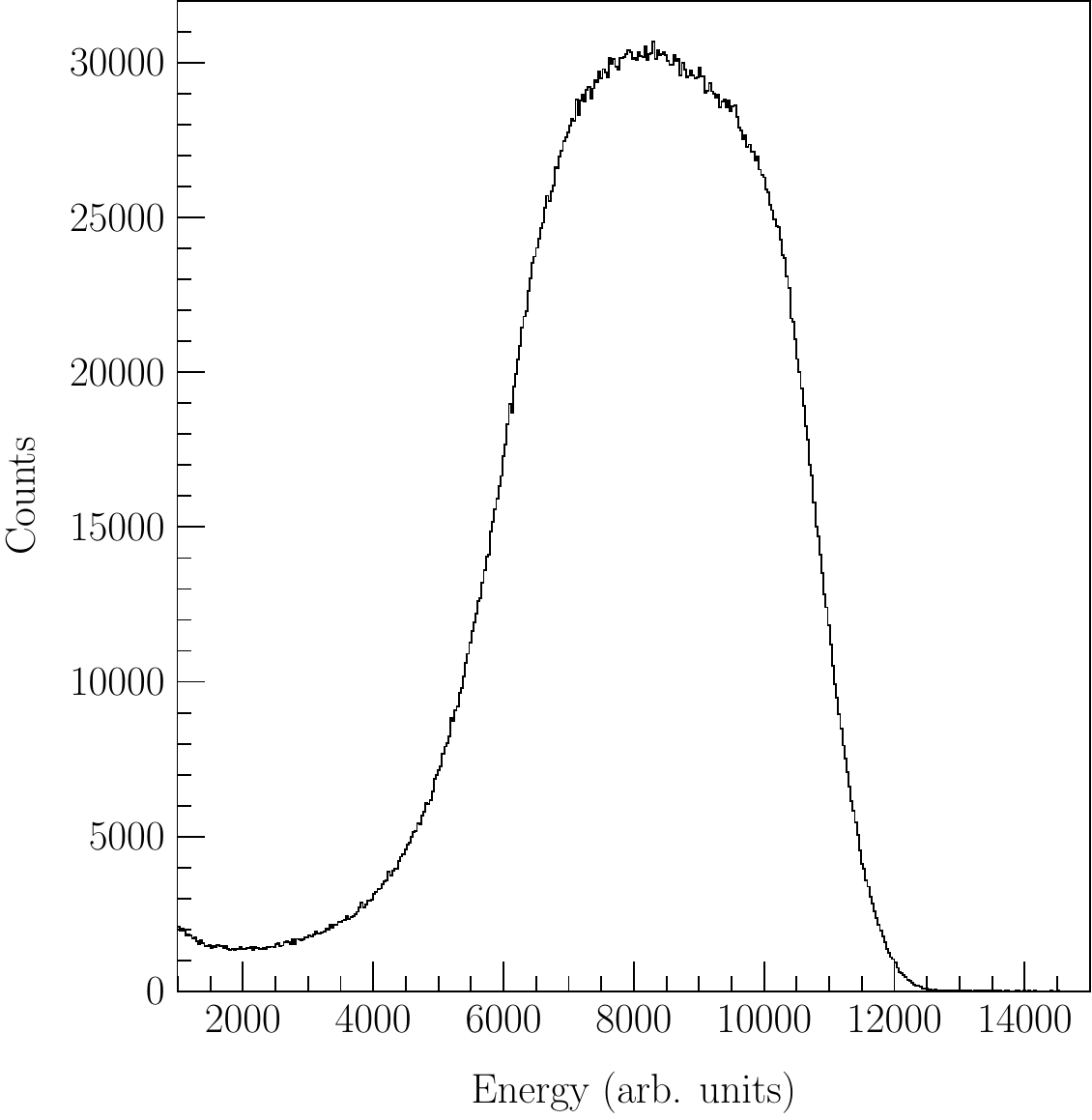}
	\caption{
		A representative photodiode energy spectrum from the \nuc{58}{Fe} measurement. The peak shape corresponds to that of Rutherford backscattering of the \nuc{58}{Fe} beam nuclei from the $^{197}$Au target.\label{fig:spectrum-particles}
	}
\end{figure}

States in \nuc{56}{Fe} and \nuc{58}{Fe} were examined via particle-$\gamma$ coincidence spectroscopy using the CAESAR array of Compton-suppressed High-Purity Germanium (HPGe) detectors and an array of silicon photodiodes.
The HPGe array consisted of seven HPGe detectors, of which six were used during the \nuc{56}{Fe} measurement.
The particle-detector array inside the target chamber consists of eight rectangular silicon photodiodes, with active areas of $25.17 \times 9.25$~mm$^2$.
The photodiodes were positioned approximately symmetrically about the beam axis at backward angles in the horizontal plane; the details are given in Table~\ref{tab:caesar-pd}.
\begin{table}
	\begin{ruledtabular}
		\caption{
			Positions of the silicon photodiode particle detectors in the CAESAR array.
			The $z$-axis is defined to be the beam direction, $\theta$ is the polar angle, and $\phi$ is the azimuthal angle, where $\phi = 0$ is vertically upwards, and $\Omega$ is the solid angle.
			The angles are to the centre of the photodiodes.
			Detector `c' was faulty during the \nuc{56}{Fe} experiment, and was removed from the data in the analysis of that measurement.\label{tab:caesar-pd}
		}
		\begin{tabular}{cddd}
			Photodiode & \mc{$\theta$ $(^\circ)$} & \mc{$\phi$ $(^\circ)$} & \mc{$\Omega$ (msr)} \\
			\hline
			a          & 122                      & 90                     & 62                  \\
			b          & 133                      & 90                     & 64                  \\
			c          & 144                      & 90                     & 64                  \\
			d          & 154                      & 90                     & 98                  \\
			e          & 151                      & 270                    & 106                 \\
			f          & 141                      & 270                    & 69                  \\
			g          & 131                      & 270                    & 73                  \\
			h          & 120                      & 270                    & 63
		\end{tabular}
	\end{ruledtabular}
\end{table}
A representative photodiode energy spectrum is given in Figure~\ref{fig:spectrum-particles}.

Preamplified signals from the HPGe detectors, their associated bismuth-germanate Compton suppressors, and the photodiodes were processed by an XIA Pixie-16 digital data acquisition system.
The data were acquired in list mode, and then coincidence events were constructed within a timing window of 3~$\mu$s.
The particle-$\gamma$ time difference spectrum showed a clear, narrow prompt peak, so a timing cut was made around this peak with random background subtractions from both sides.
A $\gamma$-ray source of \nuc{152}{Eu} was used for energy and detection-efficiency calibrations.
Doppler-shift corrections were applied to $\gamma$-ray energies for each HPGe-photodiode pair based on the reaction kinematics and the detector locations.
The energy-calibrated, particle-gated $\gamma$-ray spectrum for each individual HPGe-photodiode pair was stored in a ROOT~\cite{Brun1997} histogram for analysis.

\begin{figure}
	\includegraphics[width=8.6cm]{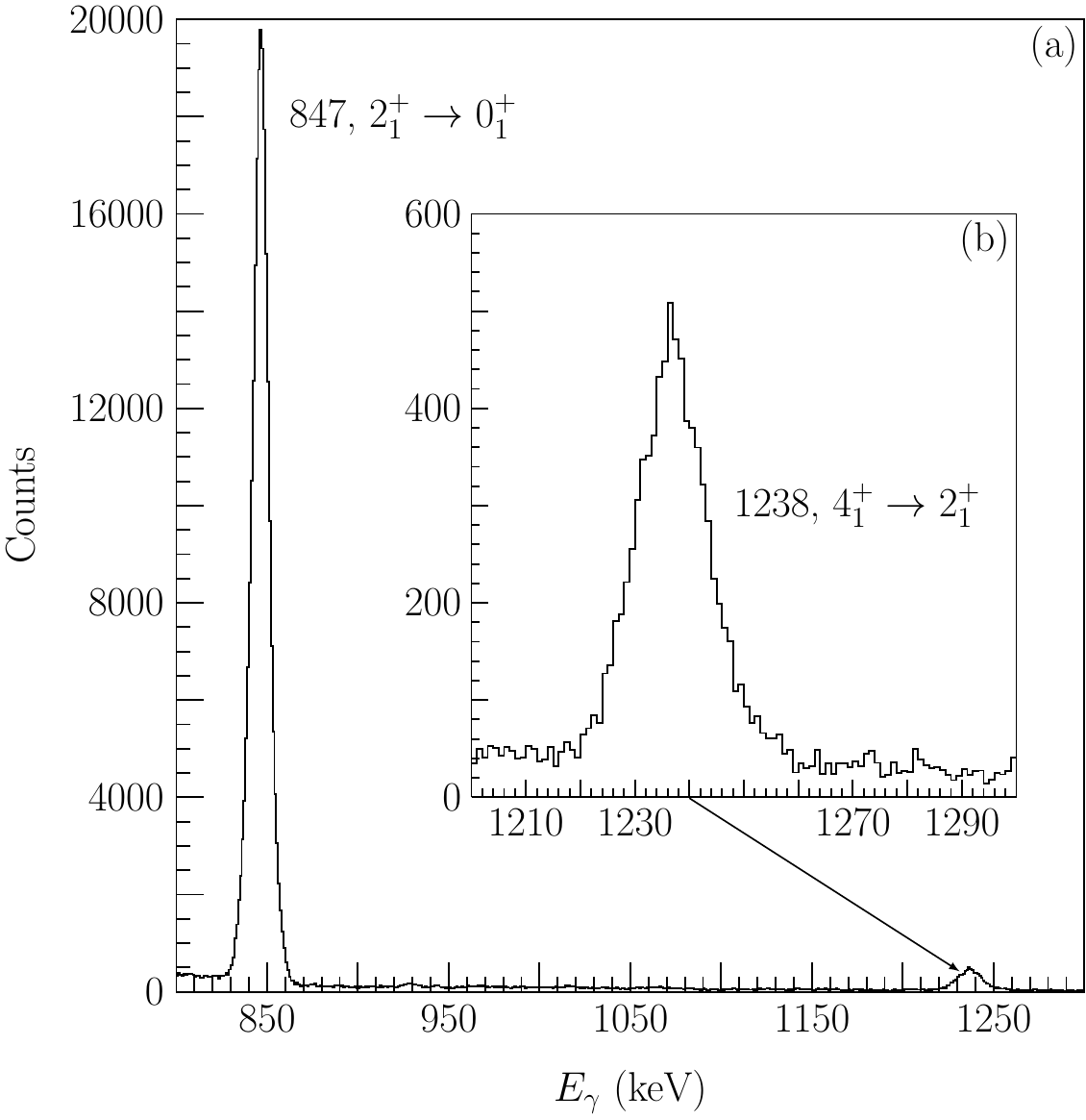}
	\caption{
	The particle-gated $\gamma$-ray spectrum collected during the measurement with the \nuc{56}{Fe} beam, summed over all detectors.
	Transitions detected following decay of Coulomb-excited \nuc{56}{Fe} are labeled by their energies in keV and $I_i^{\pi} \to I_f^{\pi}$.
	Both transitions can be seen in (a) for comparison, with (b) expanded to show the much weaker $4_1^+ \to 2_1^+$ transition.\label{fig:spectrum-56fe}
	}
\end{figure}

\begin{figure}
	\includegraphics[width=8.6cm]{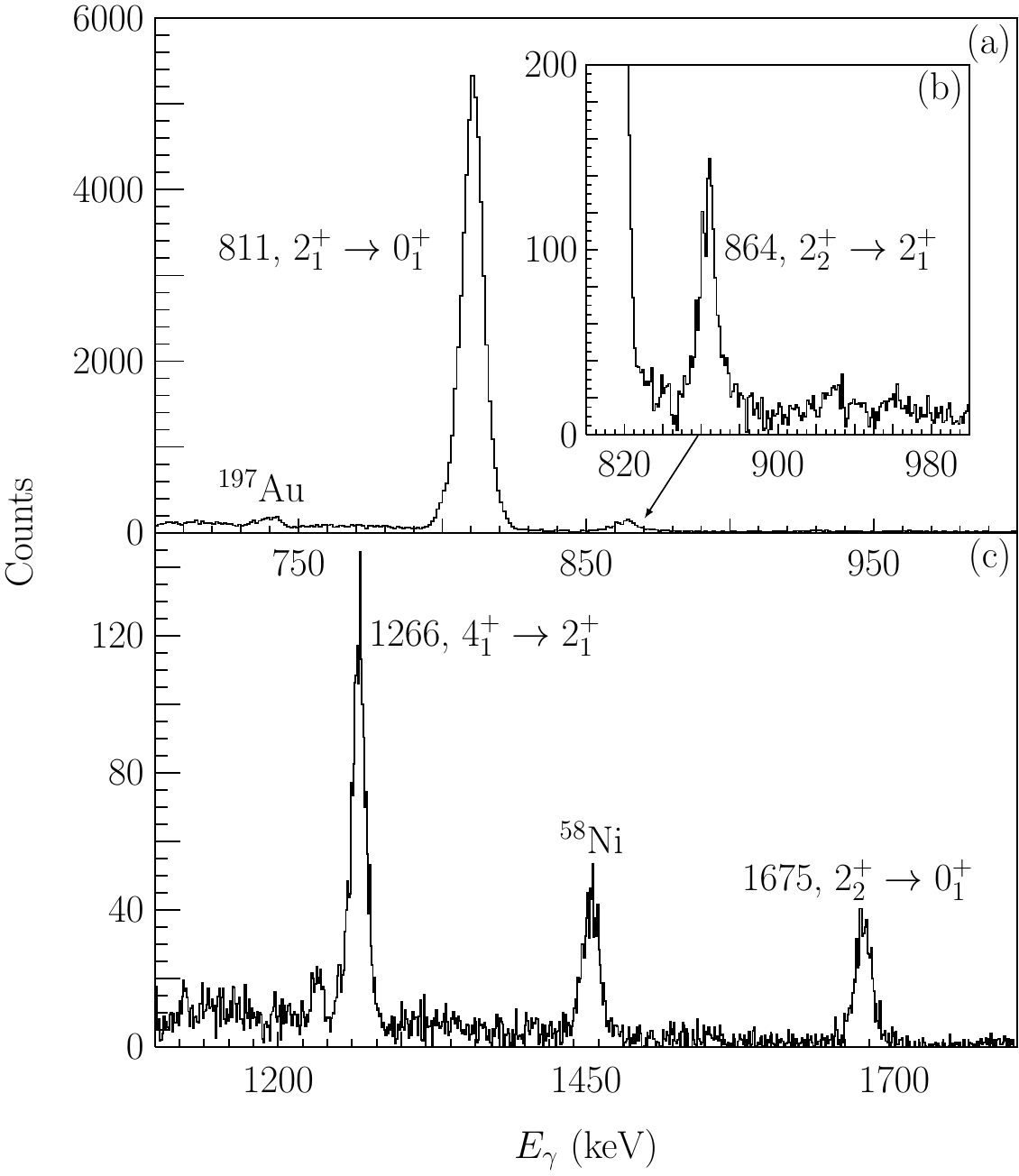}
	\caption{
	The particle-gated $\gamma$-ray spectrum collected during the measurement with the \nuc{58}{Fe} beam, summed over all detectors.
	Transitions detected following decay of Coulomb-excited \nuc{58}{Fe} are labeled, along with \nuc{58}{Ni}, which is present as a beam contaminant, and a line from the \nuc{197}{Au} target.
	Transitions from \nuc{58}{Fe} are labeled by their energies in keV and $I_i^{\pi} \to I_f^{\pi}$.
	The two lower-energy transitions can be seen in (a), along with a line from the target, with (b) expanded to show the $2_2^+ \to 2_1^+$ transition.
	The two higher-energy transitions in \nuc{58}{Fe} can be seen in (c), along with the 1454-keV $2^+_1 \rightarrow 0^+_1$ line from the \nuc{58}{Ni} beam contaminant.\label{fig:spectrum-58fe}
	}
\end{figure}

\begin{figure}
	\includegraphics[width=8.6cm]{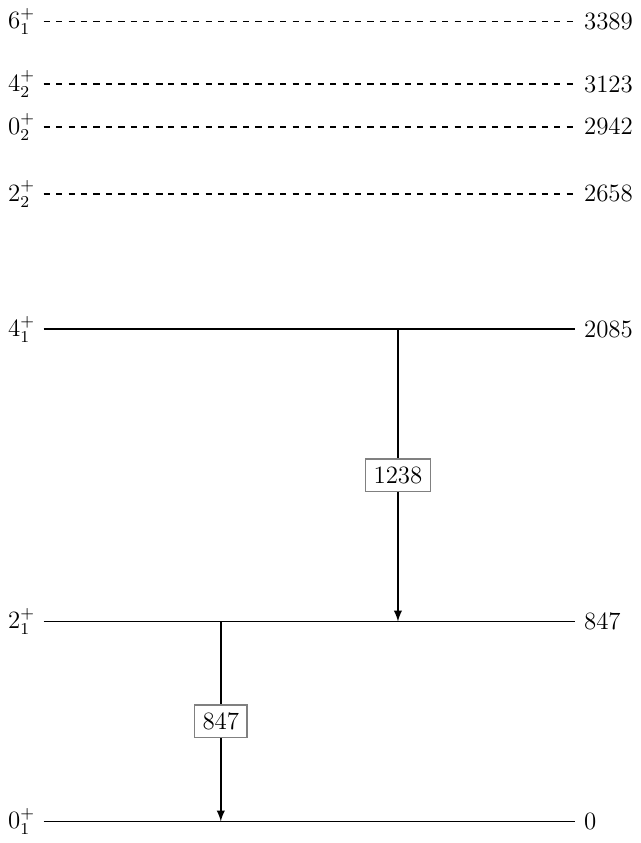}
	\caption{
		Partial level scheme of \nuc{56}{Fe}, showing the observed transitions.
		Dashed lines represent additional buffer states included in the analysis.
		Level properties and transition energies are taken from Ref.~\cite{Junde2011}, with energies given to the nearest keV.\label{fig:levels-56fe}
	}
\end{figure}

\begin{figure}
	\includegraphics[width=8.6cm]{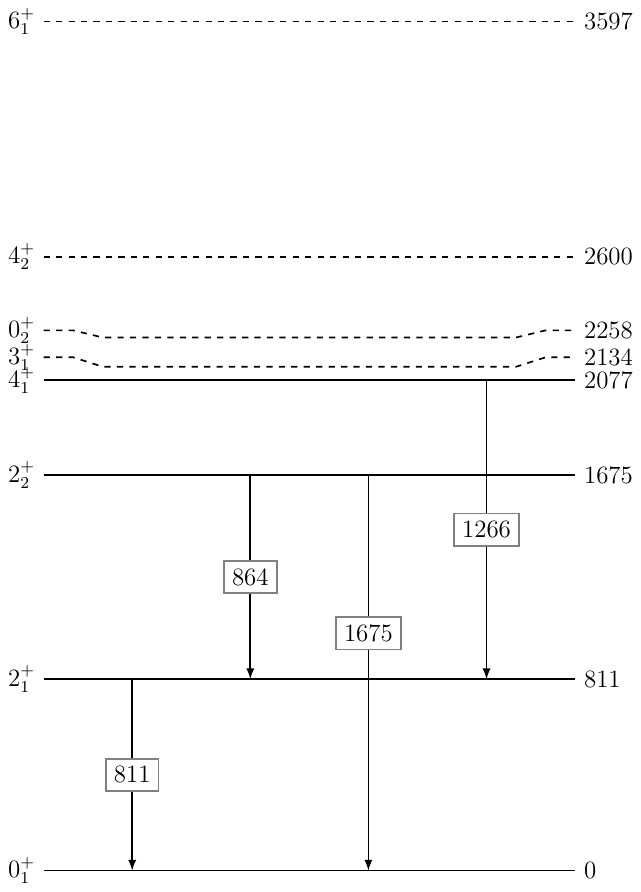}
	\caption{
		Partial level scheme of \nuc{58}{Fe}, showing the observed transitions.
		Dashed lines represent additional buffer states included in the analysis.
		Level properties and transition energies are taken from Ref.~\cite{Nesaraja2010}, with energies given to the nearest keV.\label{fig:levels-58fe}
	}
\end{figure}

The relevant sections of the total particle-gated, background-subtracted, $\gamma$-ray spectra collected with the \nuc{56}{Fe} and \nuc{58}{Fe} beams are shown in Fig.~\ref{fig:spectrum-56fe} and Fig.~\ref{fig:spectrum-58fe}, respectively.
Two transitions from the known level scheme~\cite{Junde2011} were identified in the \nuc{56}{Fe} measurement, corresponding to the decays of the 2$^+_1$ and 4$^+_1$ states; these are shown in the level scheme in Fig.~\ref{fig:levels-56fe}.
Four transitions from the adopted level scheme~\cite{Nesaraja2010} were identified in the \nuc{58}{Fe} measurement associated with the decays of the 2$^+_1$, 2$^+_2$ and 4$^+_1$ states; these are shown in the level scheme in Fig.~\ref{fig:levels-58fe}.
Note that a small feature at approximately $1232$~keV can be seen in the \nuc{58}{Fe} spectrum (Fig.~\ref{fig:spectrum-58fe}).
The origin of this peak could not be identified as either \nuc{58}{Fe} or a nearby nuclide from nucleon transfer, although it appears to be a beam-like species because the line is sharp.

Due to limited statistics, data from the individual particle-gated $\gamma$-ray spectra were summed over the HPGe detectors to produce one spectrum per photodiode.
Measured $\gamma$-ray yields were then corrected for relative HPGe detection efficiency and particle-$\gamma$ angular-correlation effects to produce an effective $4\pi$ $\gamma$-ray yield.
The observed $\gamma$-ray yields are presented in Table~\ref{tab:yields}.
\begin{table}
	\newcolumntype{s}{D{.}{.}{1.10}}
	\begin{ruledtabular}
		\caption{
			Total number of counts in the observed $\gamma$-ray transitions, and their normalized yields corrected for relative HPGe detection efficiency and particle-$\gamma$ angular-correlation effects to produce an effective $4\pi$ coverage.
			Yields are normalized to the $2_1^+ \to 0_1^+$ transition in each isotope.\label{tab:yields}
		}
		\begin{tabular}{ccss}
			Isotope      & Transition        & \mc{Counts}            & \mc{Yield}              \\
			\hline
			\nuc{56}{Fe} & $2_1^+ \to 0_1^+$ & 2.250(5) \times{} 10^5 & 1                       \\
			             & $4_1^+ \to 2_1^+$ & 7.1(1) \times{} 10^3   & 4.4(1) \times{} 10^{-2} \\
			\hline
			\nuc{58}{Fe} & $2_1^+ \to 0_1^+$ & 5.07(2) \times{} 10^4  & 1                       \\
			             & $4_1^+ \to 2_1^+$ & 1.68(5) \times{} 10^3  & 5.4(2) \times{} 10^{-2} \\
			             & $2_2^+ \to 2_1^+$ & 1.33(5) \times{} 10^3  & 3.8(1) \times{} 10^{-2} \\
			             & $2_2^+ \to 0_1^+$ & 6.8(3) \times{} 10^2   & 2.4(1) \times{} 10^{-2}
		\end{tabular}
	\end{ruledtabular}
\end{table}

\section{Analysis\label{sec:analysis}}
\subsection{Coulomb Excitation\label{sec:analysis-coulex}}
The semi-classical Coulomb-excitation code \textsc{Gosia}~\cite{Czosnyka1982} was used to extract reduced $E2$ matrix elements from the measured $\gamma$-ray yields.
Stopping powers for the \textsc{Gosia} inputs were calculated using the code SRIM~\cite{Ziegler2010}, and internal conversion coefficients were calculated using the code \textsc{BrIcc}~\cite{Kibedi2008}.

In \textsc{Gosia}, the relative yields were fitted by varying matrix elements relative to a reference transition.
In both isotopes, the $2_1^+ \to 0_1^+$ transition was used as the reference.
A fixed value for the reference transition must be chosen for the minimization procedure.
The reduced matrix element then can be obtained from the reduced transition strength.
The off-diagonal reduced matrix elements of the $E2$ operator are related to the reduced transition strengths by
\begin{equation}
	\rme{I_f}{M(E2)}{I_i}^2 = {\left(2I_i + 1\right)}B{\left(E2;I_i \to I_f\right)},
\end{equation}
where $I_i$ and $I_f$ are the initial and final states, $M(E2)$ is the $E2$ operator, and $\rme{I_f}{M(E2)}{I_i}$ denotes the reduced matrix element corresponding to the transition $I_i \to I_f$.
Note that when using an experimental $B(E2)$ value to obtain a value for the reduced matrix element, this expression gives only the magnitude of the matrix element, not its phase.

For \nuc{56}{Fe}, a reduced transition strength of $B{\left(E2; 2_1^+ \to 0_1^+\right)} = 16.8(7)$~W.u.~\cite{Junde2011} was used to obtain the fixed matrix element $\lvert\rme{0_1^+}{M(E2)}{2_1^+}\rvert = 0.327(7)$~$e$b.
For \nuc{58}{Fe}, a value of $B{\left(E2; 2_1^+ \to 0_1^+\right)} = 18.5(6)$~W.u.~\cite{Nesaraja2010} was used to obtain the fixed matrix element $\lvert\rme{0_1^+}{M(E2)}{2_1^+}\rvert = 0.351(6)$~$e$b.
Adjusting these values to the limits of the experimental uncertainties was found to change the fitted matrix elements by less than $1\%$.

For the \nuc{56}{Fe} analysis, matrix elements for the buffer states (see Fig.~\ref{fig:levels-56fe}) were fixed at values derived from the adopted spectroscopic data in ENSDF~\cite{Junde2011}.
For the \nuc{58}{Fe} analysis, matrix elements for the buffer states (see Fig.~\ref{fig:levels-58fe}) were fixed at values derived using a combination of the re-evaluated lifetimes (see Section~\ref{sec:analysis-dsam} for further details) and adopted spectroscopic data in ENSDF~\cite{Nesaraja2010}.
Note that the relative phases of the matrix elements can have a significant impact on the Coulomb-excitation yields, but individual phases are not observables as only \textit{relative} phases are important.
The phase of each level's wavefunction was chosen by fixing a sign for one matrix element to each state, and then allowing the remaining matrix elements to vary relative to these fixed signs.
Summaries of the fixed matrix elements employed are given in Tables~\ref{tab:me-56fe} and~\ref{tab:me-58fe} for \nuc{56}{Fe} and \nuc{58}{Fe}, respectively.
\begin{table}
	\newcolumntype{i}{D{x}{}{-1}}
	\begin{ruledtabular}
		\caption{
			Matrix elements that were fixed during the \nuc{56}{Fe} \textsc{Gosia} analysis.
			Unless otherwise stated, the values are derived from the adopted lifetimes, energies, branching ratios, and $E2/M1$ mixing ratios from ENSDF~\cite{Junde2011}.
			Note that the fitted $\rme{4_1^+}{M(E2)}{2_1^+}$ matrix element was fixed as positive to match the yrast band.\label{tab:me-56fe}
		}
		\begin{tabular}{cd}
			\multicolumn{2}{c}{$E2$ Matrix Elements}             \\
			\hline
			Matrix Element              & \mc{Value ($e$b)}      \\
			\hline
			$\rme{2_1^+}{M(E2)}{0_1^+}$ & +0.327\footnotemark[1] \\
			$\rme{2_2^+}{M(E2)}{0_1^+}$ & +0.055                 \\
			$\rme{2_2^+}{M(E2)}{2_1^+}$ & \pm0.145               \\
			$\rme{0_2^+}{M(E2)}{2_1^+}$ & +0.055                 \\
			$\rme{4_2^+}{M(E2)}{2_1^+}$ & +0.039                 \\
			$\rme{6_1^+}{M(E2)}{4_1^+}$ & +0.258                 \\
			$\rme{6_1^+}{M(E2)}{4_2^+}$ & \pm1.573               \\
			\hline
			\multicolumn{2}{c}{$M1$ Matrix Elements}             \\
			\hline
			                            & \mc{Value ($\mu_N^2$)} \\
			\hline
			$\rme{2_2^+}{M(M1)}{2_1^+}$ & \mp1.219               \\
			$\rme{4_2^+}{M(M1)}{4_1^+}$ & \pm2.601               \\
		\end{tabular}
	\end{ruledtabular}
	\footnotetext[1]{Weighted average of previous Coulomb-excitation measurements~\cite{Sprouse1969,Cameron1972,LeVine1981}.}
\end{table}
\begin{table}
	\newcolumntype{i}{D{x}{}{-1}}
	\begin{ruledtabular}
		\caption{
			Matrix elements that were fixed during the \nuc{58}{Fe} \textsc{Gosia} analysis.
			Unless otherwise stated, the values are derived from the re-evaluated lifetimes (Section~\ref{sec:analysis-dsam}), along with the energies, branching ratios, and $E2/M1$ mixing ratios given in ENSDF~\cite{Nesaraja2010}.
			Note that the fitted $\rme{4_1^+}{M(E2)}{2_1^+}$ matrix element was fixed as positive to match the yrast band, and $\rme{2_2^+}{M(E2)}{0_1^+}$ was fixed as positive for the $2_2^+$ state.\label{tab:me-58fe}
		}
		\begin{tabular}{cd}
			\multicolumn{2}{c}{$E2$ Matrix Elements}             \\
			\hline
			Matrix Element              & \mc{Value ($e$b)}      \\
			\hline
			$\rme{2_1^+}{M(E2)}{0_1^+}$ & +0.351\footnotemark[1] \\
			$\rme{3_1^+}{M(E2)}{2_1^+}$ & +0.063                 \\
			$\rme{0_2^+}{M(E2)}{2_1^+}$ & +0.057\footnotemark[2] \\
			$\rme{4_2^+}{M(E2)}{2_1^+}$ & \pm0.094               \\
			$\rme{4_2^+}{M(E2)}{2_2^+}$ & +0.376                 \\
			$\rme{4_2^+}{M(E2)}{4_1^+}$ & \pm0.344               \\
			$\rme{6_1^+}{M(E2)}{4_1^+}$ & +0.683                 \\
			\hline
			\multicolumn{2}{c}{$M1$ Matrix Elements}             \\
			\hline
			                            & \mc{Value ($\mu_N^2$)} \\
			\hline
			$\rme{3_1^+}{M(M1)}{2_1^+}$ & -0.174                 \\
			$\rme{3_1^+}{M(M1)}{2_2^+}$ & +0.550                 \\ 
			$\rme{4_2^+}{M(M1)}{4_1^+}$ & \mp1.002               \\
			$\rme{4_2^+}{M(M1)}{3_1^+}$ & +0.706                 \\
		\end{tabular}
	\end{ruledtabular}
	\footnotetext[1]{Previous Coulomb-excitation measurement~\cite{LeVine1981}.}
	\footnotetext[2]{Largest magnitude from the $0_2^+$ lifetime limit.}
\end{table}

The ENSDF adopted values for the branching ratios and $E2/M1$ mixing ratios were included as constraints in the \textsc{Gosia} minimization.
Some lifetimes from the re-evaluation performed below were included as constraints for the \textsc{Gosia} minimization.
The experimental values for these can be found in Tables~\ref{tab:spectroscopic-56fe} and~\ref{tab:spectroscopic-58fe} for \nuc{56}{Fe} and \nuc{58}{Fe}, respectively.
\begin{table}
	\newcolumntype{i}{D{x}{}{-1}}
	\begin{ruledtabular}
		\caption{
			Summary of the branching ratios and $E2/M1$ mixing ratios~\cite{Junde2011} included in the \nuc{56}{Fe} \textsc{Gosia} analysis.\label{tab:spectroscopic-56fe}
		}
		\begin{tabular}{ci}
			Decay branch                        & \mc{Branching ratio (\%)} \\
			\hline
			$2_2^+ \to 0_1^+ / 2_2^+ \to 2_1^+$ & 3x.1(3)                   \\
			r
			$4_2^+ \to 2_1^+ / 4_2^+ \to 4_1^+$ & 0x.85(5)                  \\
			$6_1^+ \to 4_2^+ / 6_1^+ \to 4_1^+$ & 1x.3(3)                   \\
			\hline
			Transition                          & \mc{Mixing ratio}         \\
			\hline
			$2_2^+ \to 2_1^+$                   & -0x.18(1)                 \\
		\end{tabular}
	\end{ruledtabular}
\end{table}
\begin{table}
	\newcolumntype{i}{D{x}{}{-1}}
	\begin{ruledtabular}
		\caption{
			Summary of the branching ratios, $E2/M1$ mixing ratios, and lifetimes included in the \nuc{58}{Fe} \textsc{Gosia} analysis.
			The branching ratios, mixing ratios, and $2_1^+$ lifetime are the adopted values from Ref.~\cite{Nesaraja2010}.
			The $2_2^+$ and $4_1^+$ lifetimes are the re-evaluated results of Ref.~\cite{Bolotin1978} from Section~\ref{sec:analysis-dsam}.\label{tab:spectroscopic-58fe}
		}
		\begin{tabular}{ci}
			Decay branch                        & \mc{Branching ratio (\%)} \\
			\hline
			$2_2^+ \to 0_1^+ / 2_2^+ \to 2_1^+$ & 76x.4(15)                 \\
			$3_1^+ \to 2_2^+ / 3_1^+ \to 2_1^+$ & 36x(2)                    \\ 
			$4_2^+ \to 2_1^+ / 4_2^+ \to 4_1^+$ & 77x.4(33)                 \\
			$4_2^+ \to 2_2^+ / 4_2^+ \to 4_1^+$ & 45x.5(22)                 \\
			$4_2^+ \to 3_1^+ / 4_2^+ \to 4_1^+$ & 34x.3(16)                 \\
			\hline
			Transition                          & \mc{Mixing ratio}         \\
			\hline
			$2_2^+ \to 2_1^+$                   & -0x.69(5)                 \\
			$3_1^+ \to 2_1^+$                   & -0x.40(5)                 \\
			$4_2^+ \to 4_1^+$                   & -0x.15(5)                 \\
			\hline
			State                               & \mc{Mean Lifetime (ps)}   \\
			\hline
			$2_1^+$                             & 9x.4(3)                   \\
			$2_2^+$                             & 2x.4(5)                   \\
			$4_1^+$                             & 0x.66(8)
		\end{tabular}
	\end{ruledtabular}
\end{table}
Note that for the \nuc{58}{Fe} analysis only the $2_1^+$, $2_2^+$, and $4_1^+$ lifetimes (i.e.\ states which we observed transitions from) were included as contraints in the minimization.
As the remaining (buffer) states were not observed, matrix elements for their transitions were fixed to values derived from their measured lifetimes.

\begin{figure*}[!ht]
	\includegraphics[width=8.6cm]{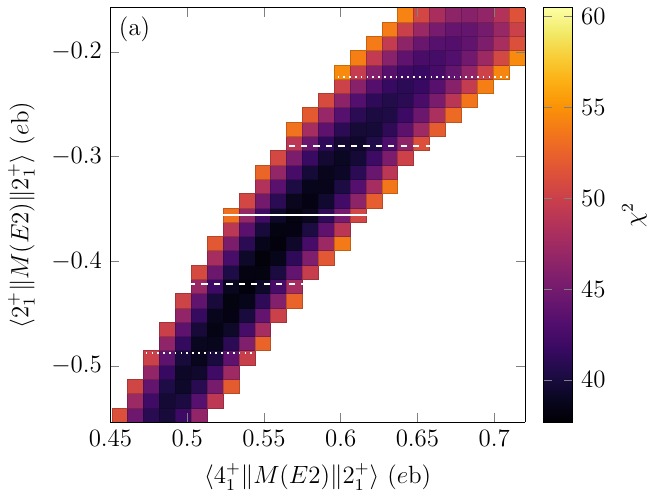}
	\includegraphics[width=8.6cm]{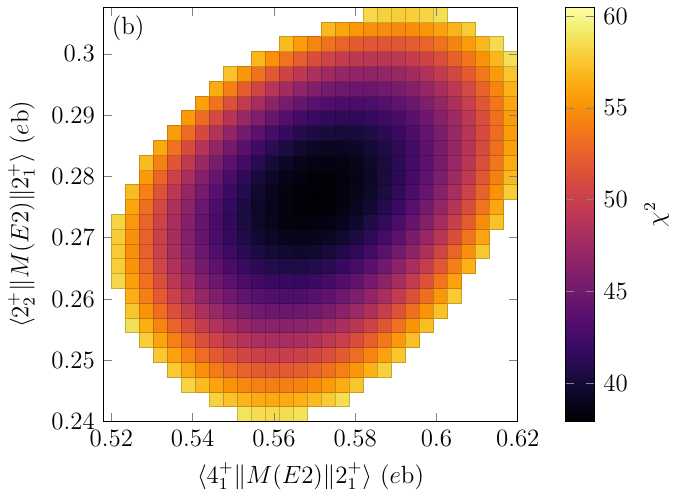}
	\caption{
		Two-dimensional $\chi^2$ surfaces for the matrix element of the $4_1^+ \to 2_1^+$ transition against the matrix elements of (a) the $2_1^+$ quadrupole moment and (b) the $2_2^+ \to 2_1^+$ transition.
		The surface is plotted up to the $95\%$ confidence limit of a $\chi^2$ distribution~\cite{bevington2003}.
		In subfigure (a), the solid horizontal line represents the adopted value from Ref.~\cite{Stone2005}, and the dashed and dotted lines represent the adopted value's $1$-$\sigma$ and $2$-$\sigma$ uncertainties, respectively.\label{fig:chisqsurfaces}
	}
\end{figure*}

The diagonal matrix element $\rme{2_1^+}{M(E2)}{2_1^+}$ was fixed in the minimization process for both isotopes.
The values were calculated from the adopted values for the spectroscopic quadrupole moments, ${Q_s(2_1^+) = -0.19(8)}$~$e$b and ${Q_s(2_1^+) = -0.27(5)}$~$e$b for \nuc{56}{Fe} and \nuc{58}{Fe}, respectively~\cite{Stone2005}.
The diagonal matrix element is related to the spectroscopic quadrupole moment by~\cite{Newton1975}
\begin{equation}
	Q_s(I) = \sqrt{\frac{16\pi}{5}}\sqrt{\frac{I (2I-1)}{(2I+1) (I+1) (2I+3)}} \rme{I}{M(E2)}{I}.
\end{equation}
The resulting matrix elements have relatively large uncertainties of $\sim 40\%$ and $\sim 20\%$, respectively.
The diagonal matrix element $\rme{2_1^+}{M(E2)}{2_1^+}$ influences the population of states connected to the $2_1^+$ state, so the uncertainties in the fitted matrix elements are dominated by the uncertainty in this value.
Two-dimensional $\chi^2$ surfaces were calculated for the $\rme{4_1^+}{M(E2)}{2_1^+}$ matrix element against the $\rme{2_1^+}{M(E2)}{2_1^+}$ and $\rme{2_2^+}{M(E2)}{2_1^+}$ matrix elements, which can be seen in Fig.~\ref{fig:chisqsurfaces}.
From this figure it can be seen that the fitted $\rme{4_1^+}{M(E2)}{2_1^+}$ matrix element is strongly correlated with the $\rme{2_1^+}{M(E2)}{2_1^+}$ diagonal matrix element.
However, the data are not able to give a well-defined minimum, i.e.\ varying one of the matrix elements does not affect the quality of the fit significantly; only the value of the other matrix element changes.
The surface for the $\rme{4_1^+}{M(E2)}{2_1^+}$ and $\rme{2_2^+}{M(E2)}{2_1^+}$ matrix elements on the other hand has a well-defined minimum.

The effect of the $\rme{4_1^+}{M(E2)}{4_1^+}$ matrix element corresponding to the $Q_s(4_1^+)$ quadrupole moment was also investigated.
Whilst the $\rme{4_1^+}{M(E2)}{2_1^+}$ value is affected by the choice of the $\rme{4_1^+}{M(E2)}{4_1^+}$ diagonal matrix element, the two values are correlated and the present data are insufficient to decouple them. Unfortunately, unlike the $Q_s(2_1^+)$ quadrupole moment, the $Q_s(4_1^+)$ quadrupole moment has not been measured previously to constrain the fit. It is hoped that the data set from Ref.~\cite{PasqualatoUnpublished} will yield new measurements of both  $Q_s(2_1^+)$ and $Q_s(4_1^+)$. In the meantime, we have used our $^{56}$Fe data set, in combination with independent measurements on $^{56}$Fe from the literature, to set the value of $\rme{4_1^+}{M(E2)}{4_1^+}$. Specifically, for \nuc{56}{Fe}, the $\rme{4_1^+}{M(E2)}{2_1^+}$ value agrees best with the previously measured values and the shell model when the $\rme{4_1^+}{M(E2)}{4_1^+}$ diagonal matrix element is small, effectively zero. Thus, this diagonal matrix element was set to zero for the analysis of both $^{56}$Fe and $^{58}$Fe.

\begin{figure}
	\includegraphics[width=8.6cm]{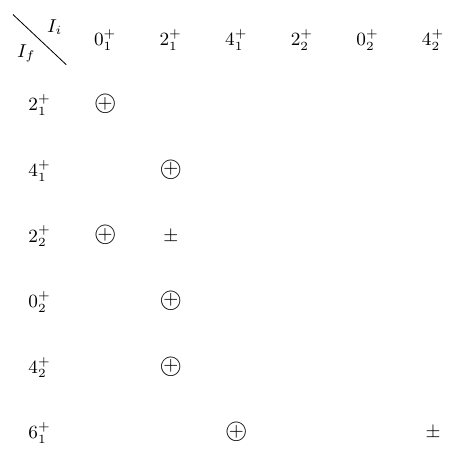}
	\caption{
		Sign combinations for the $E2$ matrix elements $\rme{I_f}{M(E2)}{I_i}$ in the \nuc{56}{Fe} analysis.
		The circled signs were chosen to fix the signs of the wavefunctions of the states, and the $\pm$ signs were varied.\label{fig:signs-56fe}
	}
\end{figure}

\begin{figure}
	\includegraphics[width=8.6cm]{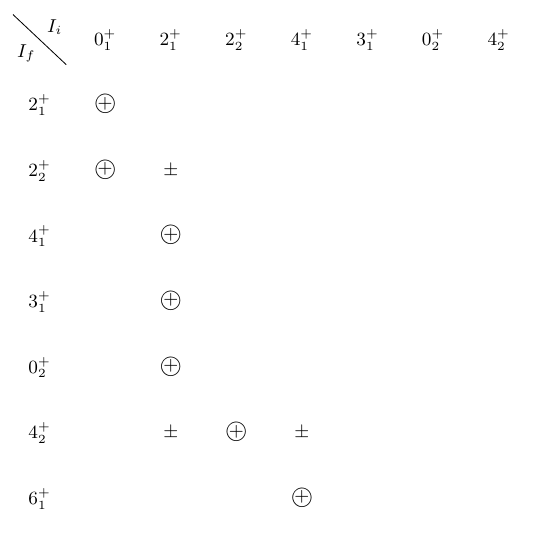}
	\caption{
		Sign combinations for the $E2$ matrix elements $\rme{I_f}{M(E2)}{I_i}$ in the \nuc{58}{Fe} analysis.
		The circled signs were chosen to fix the signs of the wavefunctions of the states, and the $\pm$ signs were varied.\label{fig:signs-58fe}
	}
\end{figure}

The relative sign combinations that were tested are shown in Fig.~\ref{fig:signs-56fe} and Fig.~\ref{fig:signs-58fe}.
For \nuc{56}{Fe} there were four sign combinations, which could not be distinguished by the data.
The sign combinations had an effect of $\approx \pm1\%$ on the fitted $\rme{4_1^+}{M(E2)}{2_1^+}$ value, which was incorporated into the uncertainty.
For \nuc{58}{Fe} there were eight sign combinations.
These combinations led to larger differences in $\chi^2$ than the \nuc{56}{Fe} minimization, and had a much larger effect on the fitted matrix elements.
In spite of this complexity, there are still some notable patterns.
Firstly, changing the sign of the $\rme{2_2^+}{M(E2)}{2_1^+}$ matrix element had a significant effect on its fitted value, resulting in a $\approx50\%$ increase compared to the opposite sign.
This change also generally resulted in a higher $\chi^2$ value, suggesting that the interference term~\cite{Alder1956}
\begin{equation}
	P_3 = {\rme{2_1^+}{M(E2)}{0_1^+}\rme{2_2^+}{M(E2)}{2_1^+}\rme{2_2^+}{M(E2)}{0_1^+}},
\end{equation}
is positive.
The positive interference term also results in much better agreement with the $B(E2)$ values derived from the re-evaluated lifetimes, those from the adopted lifetimes for the $2_2^+$ transitions, and the shell-model values.
The full set of results from the eight sign combinations is given in Table~\ref{tab:results-signs}.
\begin{table*}
	\newcolumntype{i}{D{x}{}{-1}}
	\begin{ruledtabular}
		\caption{
			Summary of the \textsc{Gosia} minimization results for the eight different sign combinations in \nuc{58}{Fe}.
			The combination is represented by the signs of $\rme{2_2^+}{M(E2)}{2_1^+}$, $\rme{4_2^+}{M(E2)}{2_1^+}$, and $\rme{4_2^+}{M(E2)}{4_1^+}$, in that order.\label{tab:results-signs}
		}
		\begin{tabular}{cdcii}
			Combination & \mc{$\chi^2$} & $I_i \to I_f$     & \mc{$\rme{I_i}{M(E2)}{I_f}$ ($e$b)} & \mc{$B(E2; I_i \to I_f)$ (W.u.)} \\
			\hline
			            &               & $4_1^+ \to 2_1^+$ & 0x.509^{+24}_{-21}                  & 22x(2)                           \\ 
			$+++$       & 46.78         & $2_2^+ \to 2_1^+$ & 0x.271^{+34}_{-29}                  & 11x(3)                           \\ 
			            &               & $2_2^+ \to 0_1^+$ & 0x.078^{+9}_{-7}                    & 0x.9(2)                          \\ 
			\hline
			            &               & $4_1^+ \to 2_1^+$ & 0x.571^{+32}_{-28}                  & 27x(3)                           \\ 
			$++-$       & 38.56         & $2_2^+ \to 2_1^+$ & 0x.278^{+35}_{-31}                  & 12x(3)                           \\ 
			            &               & $2_2^+ \to 0_1^+$ & 0x.080^{+9}_{-8}                    & 1x.0(2)                          \\ 
			\hline
			            &               & $4_1^+ \to 2_1^+$ & 0x.543^{+26}_{-24}                  & 25x(2)                           \\ 
			$+-+$       & 41.75         & $2_2^+ \to 2_1^+$ & 0x.295^{+37}_{-31}                  & 13x(3)                           \\ 
			            &               & $2_2^+ \to 0_1^+$ & 0x.084^{+10}_{-8}                   & 1x.1(3)                          \\ 
			\hline
			            &               & $4_1^+ \to 2_1^+$ & 0x.545^{+30}_{-27}                  & 25x(3)                           \\ 
			$+--$       & 42.74         & $2_2^+ \to 2_1^+$ & 0x.296^{+36}_{-31}                  & 13x(3)                           \\ 
			            &               & $2_2^+ \to 0_1^+$ & 0x.085^{+10}_{-8}                   & 1x.0(3)                          \\ 
			\hline
			            &               & $4_1^+ \to 2_1^+$ & 0x.474^{+22}_{-19}                  & 19x(2)                           \\ 
			$-++$       & 70.57         & $2_2^+ \to 2_1^+$ & -0x.368^{+10}_{-11}                 & 20x(1)                           \\ 
			            &               & $2_2^+ \to 0_1^+$ & 0x.101^{+4}_{-3}                    & 1x.5(1)                          \\ 
			\hline
			            &               & $4_1^+ \to 2_1^+$ & 0x.531^{+23}_{-22}                  & 24x(2)                           \\ 
			$-+-$       & 46.69         & $2_2^+ \to 2_1^+$ & -0x.347^{+9}_{-11}                  & 18x(1)                           \\ 
			            &               & $2_2^+ \to 0_1^+$ & 0x.098^{+4}_{-2}                    & 1x.4(1)                          \\ 
			\hline
			            &               & $4_1^+ \to 2_1^+$ & 0x.505^{+23}_{-20}                  & 21x(2)                           \\ 
			$--+$       & 53.24         & $2_2^+ \to 2_1^+$ & -0x.339^{+10}_{-11}                 & 17x(1)                           \\ 
			            &               & $2_2^+ \to 0_1^+$ & 0x.095^{+3}_{-3}                    & 1x.4(1)                          \\ 
			\hline
			            &               & $4_1^+ \to 2_1^+$ & 0x.502^{+22}_{-20}                  & 21x(2)                           \\ 
			$---$       & 51.94         & $2_2^+ \to 2_1^+$ & -0x.319^{+9}_{-10}                  & 15x(1)                           \\ 
			            &               & $2_2^+ \to 0_1^+$ & 0x.090^{+3}_{-2}                    & 1x.2(1)                          
		\end{tabular}
	\end{ruledtabular}
\end{table*}
Note that the largest $B(E2; 4_1^+ \to 2_1^+)$ value from the sign combinations in Table~\ref{tab:results-signs} is $27(3)$~W.u., which is significantly lower than the ENSDF adopted value of $47(7)$~W.u., but in good agreement with the shell-model value and the re-evaluated lifetime measurement described in Section~\ref{sec:analysis-dsam}.

\begin{figure*}
	\includegraphics[width=17.2cm]{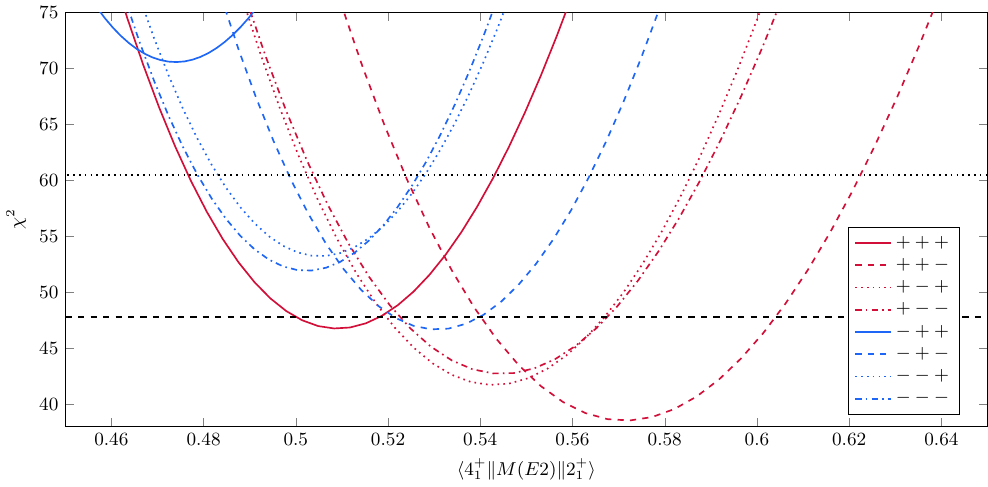}
	\caption{
		The $\chi^2$ curve for the matrix element of the $4_1^+ \to 2_1^+$ transition for each of the eight possible sign combinations.
		The combination is represented by the signs of $\rme{2_2^+}{M(E2)}{2_1^+}$, $\rme{4_2^+}{M(E2)}{2_1^+}$, and $\rme{4_2^+}{M(E2)}{4_1^+}$, in that order.
		The dashed and dotted horizontal lines respectively show the $68\%$ and $95\%$ confidence limits of the relevant $\chi^2$ distribution~\cite{bevington2003}.
		The adopted ${B(E2; 4_1^+ \to 2_1^+)}$ value in ENSDF~\cite{Nesaraja2010} corresponds to a matrix element of $0.751$~$e$b, which is significantly higher than the values from all eight combinations.\label{fig:chisqphases}
	}
\end{figure*}

The experimentally determined matrix element for \nuc{56}{Fe} was ${\rme{4_1^+}{M(E2)}{2_1^+} = 0.509^{+45}_{-42}}$~$e$b.
This corresponds to a reduced transition strength of ${B(E2; 4_1^+ \to 2_1^+) = 23(4)}$~W.u., which is in good agreement with both the ENSDF adopted value of $24(5)$~W.u., and the shell-model value of $25$~W.u.\ from our calculations.
For the case of \nuc{58}{Fe}, the one-dimensional $\chi^2$ curves of the $\rme{4_1^+}{M(E2)}{2_1^+}$ matrix element for the eight sign combinations are plotted in Fig.~\ref{fig:chisqphases}.
Although only the $-++$ combination can be excluded, as it lies above the 95\% confidence limit~\cite{bevington2003}, and the remaining seven combinations cover a wide range, the largest value is still significantly below the previously adopted value of $47(7)$~W.u., but in good agreement with the shell-model value.

\subsection{DSAM Lifetime Measurements Re-examined\label{sec:analysis-dsam}}
In this section, the discrepancy between the new Coulomb-excitation $E2$ strength and that implied by the previous lifetime measurements will be examined.
As noted above, there have been three measurements of the $4^+_1$-state lifetime in \nuc{58}{Fe}: two by the Doppler Shift Attenuation Method (DSAM)~\cite{Bolotin1978,Kosyak1983} and one by the Doppler Broadened Line Shape (DBLS) method~\cite{Cavallaro1977}.
The measurements of Cavallaro \textit{et al.}~\cite{Cavallaro1977} and Kosyak \textit{et al.}~\cite{Kosyak1983}, $\tau = 0.34(10)$ and $0.34(6)$~ps, agree well with each other, whereas the lifetime reported by Bolotin \textit{et al.}~\cite{Bolotin1978}, $\tau = 0.54^{+0.09}_{-0.07}$, is somewhat longer, although the 1-$\sigma$ uncertainties almost overlap.
Cavallaro \textit{et al.} and Bolotin \textit{et al.} used the \nuc{55}{Mn}$(\alpha,p\gamma)$\nuc{58}{Fe} reaction, whereas Kosyak \textit{et al.} used $(n,n^{\prime}\gamma)$.
An important difference between the measurements following the \nuc{55}{Mn}$(\alpha,p\gamma)$\nuc{58}{Fe} reaction is that Bolotin \textit{et al.} performed proton-$\gamma$ coincidence measurements, whereas Cavallaro \textit{et al.} did not.
By gating on the appropriate proton group, Bolotin \textit{et al.} could ensure that the level of interest is populated promptly by the reaction, and hence that feeding from higher states is eliminated.
Although the original raw data are no longer available for the measurement of Bolotin \textit{et al.}, we have access to sufficient data from the logbooks and the computer code used for the analysis.
It has been possible, therefore, to re-examine the analysis of those data, giving particular attention to the stopping powers.

The following analysis makes use of results reported in Table 2 of Bolotin \textit{et al.}, supplemented with information from the original analysis logbooks.
The computer program used by Bolotin \textit{et al.} was reconstructed based on optical character recognition of a printed listing.
After checking that the reconstructed code reproduced the original analysis, changes were made to examine the effect of modifying the stopping powers, and to automate the procedure for evaluating the level lifetimes from the observed Doppler shifts.
Some approximations employed in the original analysis were avoided.
In particular, the $F(\tau)$ function was evaluated for each excited state, thus taking into account small changes in the \nuc{58}{Fe} recoil velocity.
The original analysis used a single $F(\tau)$ curve for all states due to limited computational resources.

The experimental method employed by Bolotin \textit{et al.} has been described by Linard \textit{et al.}~\cite{Linard1978}, based on an experimental methodology first employed by Herschberger \textit{et al.}~\cite{Hershberger1969}.
Two targets, a `half' target upstream and a `full' target downstream, were placed equidistant from a {Ge(Li)} $\gamma$-ray detector midway between the two targets and facing toward the beam axis. 
The $\alpha$-particle beam energy was 10~MeV, and the targets consisted of 0.18~mg/cm$^2$ of $^{55}$Mn (i.e.\ natural Mn) evaporated onto thick Ta backings.
The incident $\alpha$-particle beam partially intercepted the upstream half-target while the remainder of the beam proceeded to the full target downstream.
Approximately equal beam intensities were maintained on the half and full targets.
The average angle of the emitted $\gamma$ rays incident upon the Ge(Li) detector was 26$^{\circ}$ to the beam axis for the upstream target and 154$^{\circ}$ for the downstream target. 
Separate annular particle detectors were positioned upstream from each target to register the reaction protons emitted in the angular range from 153$^{\circ}$ to 174$^{\circ}$ to the beam direction.
By use of this arrangement, the Doppler shift is effectively doubled compared to conventional measurements with a single particle-$\gamma$ detector combination.

\begin{figure}
	\includegraphics[width=8.6cm]{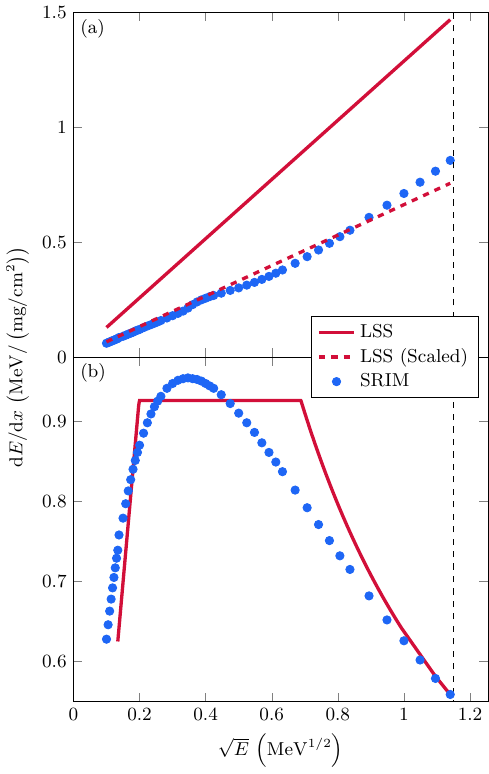}
	\includegraphics[width=8.6cm]{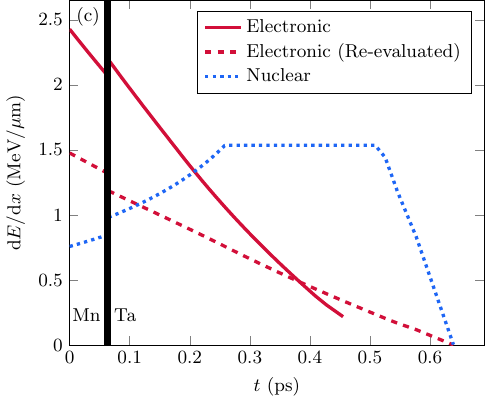}
	\caption{
		(a) Electronic and (b) nuclear stopping powers as a function of energy for \nuc{58}{Fe} in \nuc{181}{Ta} given by LSS theory used in the original analysis, SRIM calculations used in the re-evaluation, and the scaled electronic stopping powers used for the preliminary analysis.
		The dashed vertical line represents the initial velocity of the \nuc{58}{Fe} ions.
		(c) LSS stopping powers for \nuc{58}{Fe} ions passing through the target as a function of time.
		Note the discontinuity arises from the transition between the \nuc{55}{Mn} target and \nuc{181}{Ta} backing.\label{fig:stopping-powers}
	}
\end{figure}

The first step was to compare the LSS stopping powers employed in the analysis of Bolotin \textit{et al.} with current values given by SRIM~\cite{Ziegler2010}, as illustrated in Fig.~\ref{fig:stopping-powers}. 
The initial velocity is $\beta = v/c \approx 0.007$, or $v \approx 0.96 v_0$, where $v_0 = c/137$ is the Bohr velocity.
This velocity is equivalent to approximately $1.3$~MeV.
The nuclear stopping powers given by SRIM agree reasonably well with those used by Bolotin \textit{et al.} based on the parametrization of Wozniak, Hershberger and Donahue~\cite{Wozniak1969}.
Multiple scattering was handled using the formulation of Blaugrund~\cite{Blaugrund1966}.
The Lindhard, Scharff and Schi{\o}tt (LSS)~\cite{Lindhard1963} electronic stopping powers used for the lifetime analysis, however, are about twice as large as those given by SRIM, as seen in Fig.~\ref{fig:stopping-powers}.
According to SRIM, the nuclear and electronic stopping powers for \nuc{58}{Fe} in \nuc{181}{Ta} become comparable when the \nuc{58}{Fe} energy falls below approximately $0.9$~MeV, below which the nuclear stopping dominates.

Fig.~\ref{fig:stopping-powers}c illustrates how the stopping powers change with time as the $^{58}$Fe ions slow in the target.
The LSS stopping powers were used for this schematic.
It is evident that the electronic-stopping-power difference has a more pronounced effect on the Doppler shift for shorter-lived states, whereas for longer-lived states the nuclear stopping powers begin to dominate.
Bolotin \textit{et al.} used the unmodified LSS stopping powers.
The correction to LSS stopping powers proposed by Robertson~\cite{Robertson1977} would increase the LSS stopping powers higher still than those of SRIM. 

Table~\ref{tab:results-dsam} presents a re-analysis of the data of Bolotin \textit{et al.} with SRIM stopping powers.
\begin{table*}
	\begin{ruledtabular}
		\caption{
			Re-evaluated DSAM results from Bolotin \textit{et al.}~\cite{Bolotin1978} using updated stopping powers as described in the text. The experimental $F(\tau) = \Delta E_{\gamma} /(E_{\gamma} \times \Delta \langle \beta \cos \phi \rangle )$ where $\phi$ represents the angle between the direction of the recoiling nucleus and the direction of $\gamma$-ray emission~\cite{Linard1978}. The first uncertainty in $\tau$ is statistical, whereas the second includes an estimate of the uncertainty in the stopping powers, $\pm 5\%$ on electronic and $\pm 10\%$ on nuclear stopping.
			A summary is given in Table~\ref{tab:summary-dsam}, along with a comparison to previous works.
			Excitation energies $E_x$ and $J^\pi$ assignments have been updated from those in Ref.~\cite{Bolotin1978} to the current adopted values in Ref.~\cite{Nesaraja2010}. The transition energy $E_{\gamma}$ corresponds to the energy difference between the energies of the initial and final states. Both $E_x$ and $E_{\gamma}$ are rounded to the nearest keV.
			Where more than one transition depopulates a state, average $F(\tau)$ values are denoted in angle brackets.
			\label{tab:results-dsam}
		}
		\begin{tabular}{dcdddddd}
			\mc{$E_x$} & $J^{\pi}$ & \mc{$E_{\gamma}$} & \mc{$\Delta E_{\gamma}$} & \mc{$\Delta \langle \beta \cos \phi  \rangle \times 10^{3}$} & \mc{$F(\tau)$}                                & \multicolumn{2}{c}{$\tau$~(ps)}                                \\
			\cmidrule(lr){7-8}
			\mc{(MeV)} &           & \mc{(MeV)}        & \mc{(keV)}               &                                                              &                                               & \mc{Re-evalation}               & \mc{Ref.~\cite{Bolotin1978}} \\
			\hline
			0.811      & $2^+$     & 0.811             & 0.46(17)                 & 12.57                                                        & 0.045(17)                                     & >3                              & 3.4^{+1.0}_{-0.9}            \\
			1.675      & $2^+$     & 0.864             & 0.62(13)                 & 12.36                                                        & 0.058(12)                                     &                                 &                              \\
			           &           & 1.675             & 1.71(24)                 &                                                              & 0.083(12)                                     &                                 &                              \\
			\cmidrule(lr){6-6}
			           &           &                   &                          &                                                              & \langle{} 0.070(8) \rangle{}                  & 2.4(3)(5)                       & 2.30^{+0.64}_{-0.57}         \\ 
			2.077      & $4^+$     & 1.266             & 3.55(14)                 & 12.26                                                        & 0.229(9)                                      & 0.66(3)(8)                      & 0.54^{+0.09}_{-0.07}         \\  
			2.134      & $3^+$     & 0.459             & 0.16(35)                 & 12.24                                                        & 0.028(62)                                     &                                 &                              \\
			           &           & 1.323             & 0.79(18)                 &                                                              & 0.049(11)                                     &                                 &                              \\
			\cmidrule(lr){6-6}
			           &           &                   &                          &                                                              & \langle{} 0.048(11) \rangle{}                 & 3.6^{(+11)(14)}_{(-7)(10)}      & 3.15^{+1.00}_{-0.95}         \\ 
			2.258      & $0^+$     & 1.447             & 0.25(43)                 & 12.21                                                        & 0.014(24)                                     & >4                              & > 3.7                        \\
			2.600      & $4^+$     & 0.467             & 0.85(42)                 & 12.12                                                        & 0.15(7)                                       &                                 &                              \\
			           &           & 0.524             & 0.73(13)                 &                                                              & 0.115(20)                                     &                                 &                              \\
			           &           & 0.926             & 1.08(21)                 &                                                              & 0.096(19)                                     &                                 &                              \\
			           &           & 1.790             & 3.12(35)                 &                                                              & 0.144(16)                                     &                                 &                              \\
			\cmidrule(lr){6-6}
			           &           &                   &                          &                                                              & \langle{} 0.122(10) \rangle{}\footnotemark[1] & 1.36(14)(24)                    & 1.05^{+0.20}_{-0.18}         \\ 
			2.782      & $1^+$     & 0.524             & 2.55(56)                 & 12.07                                                        & 0.40(9)                                       &                                 &                              \\
			           &           & 1.107             & 4.67(53)                 &                                                              & 0.35(4)                                       &                                 &                              \\
			           &           & 1.971             & 11.39(82)                &                                                              & 0.48(3)                                       &                                 &                              \\
			           &           & 2.782             & 15.15(150)               &                                                              & 0.45(5)                                       &                                 &                              \\
			\cmidrule(lr){6-6}
			           &           &                   &                          &                                                              & \langle{} 0.429(22) \rangle{}                 & 0.285(23)(44)                   & 0.266^{+0.047}_{-0.031}      \\ 
			2.876      & $2^+$     & 2.066             & 14.27(45)                & 12.05                                                        & 0.573(18)                                     & 0.173(11)(24)                   & 0.135^{+0.020}_{-0.019}      \\ 
			3.084      & $2^+$     & 2.273             & 23.48(36)                & 11.99                                                        & 0.862(13)                                     & 0.050(5)(9)                     & 0.036^{+0.008}_{-0.006}      \\ 
			3.233      & $2^+$     & 0.633             & 2.30(0.64)               & 11.95                                                        & 0.304(85)                                     &                                 &                              \\
			           &           & 1.157             & 4.95(49)                 &                                                              & 0.358(35)                                     &                                 &                              \\
			\cmidrule(lr){6-6}
			           &           &                   &                          &                                                              & \langle{} 0.350(33)  \rangle{}                & 0.38^{(+5)(8)}_{(-4)(6)}        & 0.31^{+0.08}_{-0.06}         \\ 
			3.450      & $(4^+)$   & 0.849             & 2.55(71)                 & 11.89                                                        & 0.252(70)                                     &                                 &                              \\
			           &           & 1.316             & 3.55(71)                 &                                                              & 0.227(45)                                     &                                 &                              \\
			\cmidrule(lr){6-6}
			           &           &                   &                          &                                                              & \langle{} 0.235(38)  \rangle{}                & 0.64^{(+15)(20)}_{(-11)(16)}    & 0.52^{+0.19}_{-0.12}         \\ 
			3.538      & $1^+$     & 1.863             & 19.45(89)                & 11.86                                                        & 0.881(40)                                     &                                 &                              \\
			           &           & 2.727             & 31.48(28)                &                                                              & 0.973(9)                                      &                                 &                              \\
			\cmidrule(lr){6-6}
			           &           &                   &                          &                                                              & \langle{} 0.969(8)  \rangle{}                 & 0.012(3)(4)                     & 0.008^{+0.004}_{-0.003}      \\ 
			3.597      & $6^+$     & 1.520             & 7.84(26)                 & 11.84                                                        & 0.435(14)                                     & 0.278(14)(35)                   & 0.220^{+0.040}_{-0.025}      \\ 
			3.630      & $2^+$     & 2.819             & 32.35(28)                & 11.83                                                        & 0.969(8)                                      & 0.012(3)(4)                     & 0.008(3)                     \\ 
			3.754      & $(4)^+$   & 1.678             & 19.18(71)                & 11.80                                                        & 0.969(36)                                     & <0.025                          & < 0.019                      \\
			3.789      & $(5^-)$   & 1.713             & 17.33(29)                & 11.79                                                        & 0.858(14)                                     & 0.051(5)(9)                     & 0.037^{+0.008}_{-0.006}      \\ 
			3.880      & $1^+$     & 3.880             & 48.1(34)                 & 11.76                                                        & 1.054(75)                                     & <0.009                          & <0.005                       \\
			3.886      & $6^+$     & 0.289             & 0.84(40)                 & 11.76                                                        & 0.25(12)                                      &                                 &                              \\
			           &           & 1.810             & 3.92(73)                 &                                                              & 0.184(34)                                     &                                 &                              \\
			\cmidrule(lr){6-6}
			           &           &                   &                          &                                                              & \langle{} 0.189(33) \rangle{}                 & 0.83^{(+20)(27)}_{(-14)(20)}    & 0.68^{+0.24}_{-0.15}         \\ 
			4.015      & $1^+$     & 3.204             & 35.81(46)                & 11.72                                                        & 0.954(12)                                     & 0.018(5)(6)                     & 0.012^{+0.005}_{-0.004}      \\ 
			4.139      & $1^+$     & 4.139             & 48.61(50)                & 11.68                                                        & 1.005(10)                                     & <0.002                          & <0.001                       \\
			4.215      & $(5^+)$   & 1.614             & 3.73(49)                 & 11.66                                                        & 0.198(26)                                     & 0.78^{(+14)(20)}_{(-11)(16)}    & 0.65^{+0.20}_{-0.14}         \\ 
		\end{tabular}
	\end{ruledtabular}
	\footnotetext[1]{A typographical error in this value in Ref.~\cite{Bolotin1978} has been corrected.}
\end{table*}
As this required significant modification to the DSAM code, for a preliminary analysis the LSS stopping powers were simply scaled to better conform to the SRIM values.
Specifically, the LSS stopping powers for \nuc{58}{Fe} in \nuc{55}{Mn} and \nuc{181}{Ta} were multiplied by factors of 0.608 and 0.515, respectively.
The nuclear stopping powers were not changed.
After this approximate analysis, the DSAM code was modified to include SRIM stopping powers.
Generally this more rigourous inclusion of the SRIM stopping powers changed the lifetime by at most one in the last digit of the values in Table~\ref{tab:results-dsam} when compared to the scaled LSS analysis.
An overall uncertainty of $\pm5\%$ in the electronic stopping powers and $\pm10\%$ in the nuclear stopping powers was added to the uncertainties assigned to the extracted lifetimes.
This modification of the electronic stopping brings the lifetime of the $4^+$ state, $\tau = 0.66(6)$~ps, into agreement with the shortest lifetime determined by the present Coulomb-excitation measurement, $\tau = 0.6(1)$~ps.
A summary of the revised lifetimes with a comparison to previous works is given in Table~\ref{tab:summary-dsam}.
\begin{table}
	\begin{ruledtabular}
		\caption{
			Re-evaluated lifetimes from Bolotin \textit{et al.}~\cite{Bolotin1978} using updated stopping powers, in comparison to measurements by Cavallaro \textit{et al.}~\cite{Cavallaro1977} and Kosyak \textit{et al.}~\cite{Kosyak1983} where they exist.\label{tab:summary-dsam}
		}
		\begin{tabular}{dcddd}
			\mc{$E_x$} & $J^{\pi}$ & \mc{Cavallaro}           & \mc{Kosyak}       & \mc{Re-evaluated} \\
			\cmidrule(lr){3-5}
			\mc{(MeV)} &           & \multicolumn{3}{c}{(ps)}                                         \\
			\hline
			0.811      & $2^+$     & 11.7                     &                   & > 3               \\
			1.675      & $2^+$     &                          &                   & 2.4(5)            \\
			2.077      & $4^+$     & 0.34(10)                 & 0.34(6)           & 0.66(8)           \\
			2.134      & $3^+$     &                          &                   & 3.6^{+14}_{-10}   \\
			2.258      & $0^+$     &                          &                   & > 4               \\
			2.600      & $4^+$     & 0.54^{+18}_{-10}         & >0.4              & 1.36(24)          \\
			2.782      & $1^+$     &                          & 0.29^{+13}_{-7}   & 0.29(4)           \\
			2.876      & $2^+$     &                          & 0.14^{+3}_{-2}    & 0.17(2)           \\
			3.084      & $2^+$     &                          & 0.048^{+17}_{-12} & 0.050(9)          \\
			3.233      & $2^+$     &                          &                   & 0.38^{+8}_{-6}    \\
			3.450      & $(4^+)$   &                          &                   & 0.64^{+20}_{-16}  \\
			3.538      & $1^+$     &                          &                   & 0.012(4)          \\
			3.597      & $6^+$     & 0.49(5)                  & 0.16^{+4}_{-5}    & 0.28(3)           \\
			3.630      & $2^+$     &                          & 0.021(4)          & 0.012(4)          \\
			3.754      & $(4^+)$   &                          &                   & < 0.025           \\
			3.789      & $(5^-)$   &                          &                   & 0.051(9)          \\
			3.880      & $1^+$     &                          &                   & < 0.009           \\
			3.886      & $6^+$     & 0.70^{+22}_{-10}         &                   & 0.83^{+27}_{-20}  \\
			4.015      & $1^+$     &                          &                   & 0.018(6)          \\
			4.139      & $1^+$     &                          &                   & < 0.002           \\
			4.215      & $(5^+)$   &                          &                   & 0.78^{+20}_{-16}  \\
		\end{tabular}
	\end{ruledtabular}
\end{table}

\section{Discussion\label{sec:discussion}}
\subsection{Comparison of Experimental Data}
\begin{figure}
	\includegraphics[width=8.6cm]{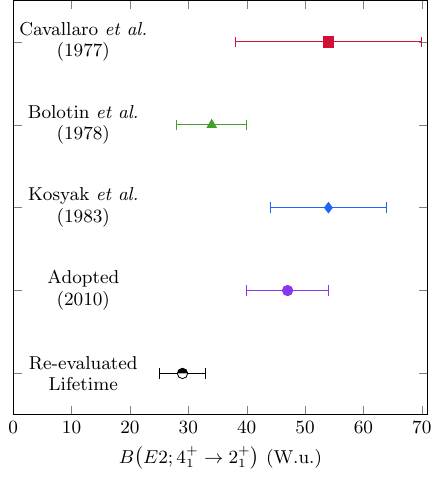}
	\caption{
		Timeline of $B{\left(E2; 4_1^+ \to 2_1^+\right)}$ values for \nuc{58}{Fe}.
		The previous measurements are from Refs.~\cite{Cavallaro1977,Bolotin1978,Kosyak1983}, and the adopted value is from Ref.~\cite{Nesaraja2010}.\label{fig:discussion-timeline}
	}
\end{figure}

There are some inconsistencies between the previous lifetime measurements in \nuc{58}{Fe} that were used to derive the $B{\left(E2; 4_1^+ \to 2_1^+\right)}$ value, as shown in Fig.~\ref{fig:discussion-timeline}.
The change in the electronic stopping powers to those of SRIM has the most effect on the shorter-lived states, which decay while the slowing is dominated by the electronic stopping.
For the longer-lived states ($\tau > 1$~ps), the energy loss, and hence the extracted lifetime, is dominated by the nuclear stopping process.
Hence, the revised lifetimes are up to 1.6 times longer for the shorter-lived states, whereas the lifetimes for the longer-lived states change little.
As seen in Table~\ref{tab:summary-dsam}, the revised lifetime results agree with those of Kosyak \textit{et al.}, apart from the $4^+_1$ lifetime, and also to some extent the $6^+_1$ lifetime.
In contrast, there is no agreement with the results of Cavallaro \textit{et al.}, except for the $6_2^+$ state.
A possible explanation is that the lower states in the work of Cavallaro \textit{et al.} are affected by feeding from higher states.
The proton-$\gamma$ measurements of Bolotin \textit{et al.} are not affected by such feeding.

Some caution is warranted for the lifetimes of longer-lived states extracted from the DSAM analysis.
Firstly, taking the example of the measurement of Bolotin \textit{et al.} in Fig.~\ref{fig:stopping-powers}, the stopping time of the \nuc{58}{Fe} recoil in Ta is on the order of 0.6~ps, so the technique loses sensitivity as the lifetime increases beyond this stopping time.
Secondly, the longer-lived states are more affected by the nuclear stopping and scattering process, which has uncertainties that are not readily quantified.
Thirdly, the observed Doppler shifts are small for the states with $\tau > 1$~ps, and may be subject to systematic uncertainties on the level of the observed Doppler shift.
Unfortunately, the original data from Ref.~\cite{Bolotin1978} are not available to assess the extracted Doppler shifts.
Table~\ref{tab:summary-dsam} presents the extracted lifetimes based on the stated stopping powers and the available Doppler-shift values, but we recommend that the longer lifetimes be remeasured by methods that are appropriate for lifetimes in the picosecond range (such as the recoil-distance Doppler-shift method).

Finally, it may be noted that Kosyak \textit{et al.} measured both \nuc{56}{Fe} and \nuc{58}{Fe}, obtaining what now appears to be an anomalous lifetime for the $4_1^+$ state in \nuc{58}{Fe} but not \nuc{56}{Fe}.
LSS stopping powers were employed.
We conjecture that this difference may be attributed to their use of an oxide target for the \nuc{56}{Fe} measurement, whereas a metal target was used for the \nuc{58}{Fe} study.
In the case of the oxide, the stopping will be dominated by the oxygen; the above discussion suggests that LSS electronic stopping powers for an iron host are not reliable.

\subsection{Shell-Model Calculations}
Shell-model calculations were performed with \textsc{NuShellX}~\cite{Brown2014} using the GXPF1A interaction, the full $fp$-shell as the basis, and with effective charges of $e_\pi = 1.5$ and $e_\nu = 0.5$~\cite{Honma2004,Honma2005}.
These are the recommended effective charges for this interaction.
There has been recent work suggesting that $e_\pi = 1.33$ and $e_\nu = 0.45$ can be used universally across the $sd$ and $fp$ shells~\cite{Ogunbeku2025}, which will be discussed further below.
Calculations were also performed for $^{56}$Fe, but as the $B{\left(E2; 4_1^+ \to 2_1^+\right)}$ value we measured is consistent with the adopted value given in Ref.~\cite{Junde2011}, and as it also agrees with the GXPF1A shell-model calculations, further comparisons of theory and experiment for $^{56}$Fe are not included here.

The shell-model reduced transition strengths calculated for \nuc{58}{Fe} with GXPF1A and the standard effective charges are given in Table~\ref{tab:shell-model}, along with a comparison to those derived from the re-evaluated lifetimes.
\begin{table}
	\newcolumntype{i}{D{x}{}{-1}}
	\begin{ruledtabular}
		\caption{
			Summary of the reduced transition strengths from the shell-model calculations for this work.
			The values derived from the re-evaluated lifetimes are also given.\label{tab:shell-model}
		}
		\begin{tabular}{cccdi}
			$E_i$~(MeV)                      & $I_i^{\pi}$ & $I_f^{\pi}$ & \multicolumn{2}{c}{$B(E2; I_i \to I_f)$~(W.u.)}                                        \\
			\cmidrule(lr){4-5}
			\mc{Adopted~\cite{Nesaraja2010}} &             &             & \mc{Shell Model}                                & \mc{Re-evaluated~\cite{Bolotin1978}} \\
			\hline
			0.811                            & $2_1^+$     & $0_1^+$     & 22                                              & \mc{---}                             \\
			1.675                            & $2_2^+$     & $0_1^+$     & 1.5                                             & 0x.91(19)                            \\
			                                 &             & $2_1^+$     & 9                                               & 10x(2)                               \\ 
			2.077                            & $4_1^+$     & $2_1^+$     & 29                                              & 29x(4)                               \\ 
			2.134                            & $3_1^+$     & $2_1^+$     & 2.2                                             & 0x.4(2)                              \\
			2.258                            & $0_2^+$     & $2_1^+$     & 5                                               & <1x.8                                \\
			2.600                            & $4_2^+$     & $2_1^+$     & 2.1                                             & 0x.74(13)                            \\
			                                 &             & $2_2^+$     & 8                                               & 12x(2)                               \\ 
			                                 &             & $4_1^+$     & 3.3                                             & 10x(7)                               \\ 
			2.782                            & $1_1^+$     & $2_1^+$     & 0.2                                             & 0x.10(5)                             \\
			                                 &             & $2_2^+$     & 2.7                                             & 0x.9(3)                              \\
			3.597                            & $6_1^+$     & $4_1^+$     & 23                                              & 27x(3)                               \\ 
			3.886                            & $6_2^+$     & $4_1^+$     & 9                                               & 2x.2(7)                              \\
			                                 &             & $4_2^+$     & 1.6                                             & 1x.2(4)
		\end{tabular}
	\end{ruledtabular}
\end{table}

\begin{figure*}
	\includegraphics[width=8.6cm]{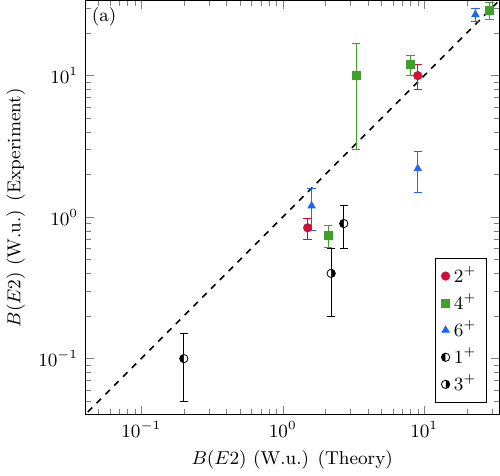}
	\includegraphics[width=8.6cm]{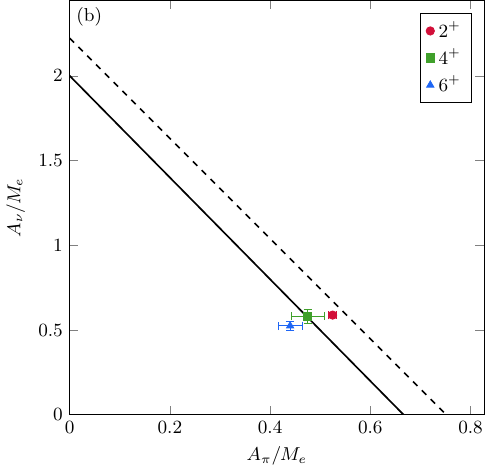}
	\caption{
		(a) Comparison of shell-model $B(E2)$ values in \nuc{58}{Fe} with those derived from the re-evaluated lifetime measurements The dotted line at $45^\circ$ represents perfect agreement. Points are marked by the spin of the initial state. 
		See Table~\ref{tab:shell-model} for numerical values and the text for further discussion.
		(b) Comparison of shell-model matrix elements to experiment for the yrast states. For experiment, the literature $\rme{2_1^+}{M(E2)}{0_1^+}$ value and the re-evaluated DSAM $\rme{4_1^+}{M(E2)}{2_1^+}$ and $\rme{6_1^+}{M(E2)}{4_1^+}$ values are adopted.
		The solid (dashed) line uses effective charges of $e_\pi = 1.5$, $e_\nu = 0.5$ ($e_\pi = 1.3$, $e_\nu = 0.45$).
		Points are again marked by the spin of the initial state.\label{fig:discussion-linear}
	}
\end{figure*}

Theory and experiment are also compared in Fig.~\ref{fig:discussion-linear}a.
With the exception of the $6^+_2 \rightarrow 4^+_1$ transition, the agreement between theory and experiment is satisfactory.
There is generally better agreement for the yrast states.
As theory and experiment agree for the $6^+_2 \rightarrow 4^+_2$ transition, it is suggested that the experimental branching ratios for the decays of the $6^+_2$ state should be checked.

Returning now to the choice of effective charges, it is useful to begin by noting that the shell-model $E2$ matrix elements $M_s$ are given by
\begin{equation}
	M_s = A_\pi e_\pi + A_\nu e_\nu,
\end{equation}
where $A_\pi$ and $A_\nu$ are the calculated transition amplitudes.
If we plot $A_\nu / M_e$ versus $A_\pi / M_e$, where $M_e$ is the experimental matrix element, the data points should fall on a straight line with slope $-e_\pi / e_\nu$ and intercepts $1/e_\nu$ at $A_\pi / M_e=0$ and $1/e_\pi$ at $A_\nu / M_e=0$.
The data for the yrast decays in $^{58}$Fe are plotted this way in Fig.~\ref{fig:discussion-linear}b, along with the lines corresponding to the two sets of effective charges.
The yrast states were selected for this comparison based on the assumption that their shell model wavefunctions might be more accurate than those of the non-yrast states.
The outcome is that the data form a cluster that does not define the slope of the line (i.e.
the ratio $-e_\pi / e_\nu$) but they better agree with the standard set of effective charges, $e_\pi = 1.5$ and $e_\nu = 0.5$ than the newly proposed ones.
More extensive calculations would be required to identify the reason for the difference.
In the interim, it can be noted that effective charges depend on the basis space and the chosen interactions.
The effective charges determined by Ogunbeku \textit{et al.} were based on calculations using the UFP-CA interaction~\cite{Magilligan2021}, which is optimized for neutron-rich calcium isotopes beyond $^{48}$Ca and is appropriate for the key neutron-rich Ca isotopes and $^{54}$Sc included in their study.
The case of $^{58}$Fe is closer to $^{56}$Ni than $^{48}$Ca and may require not only different interactions, but modified effective charges.

\section{Conclusions\label{sec:conclusions}}
Reduced transition strengths in \nuc{56,58}{Fe} have been measured by Coulomb excitation.
The $B{\left(E2; 4_1^+ \to 2_1^+\right)}$ value in \nuc{56}{Fe} and the $B{\left(E2; 2_2^+ \to 2_1^+\right)}$ and $B{\left(E2; 2_2^+ \to 0_1^+\right)}$ values in \nuc{58}{Fe} agree well with both shell-model calculations and previous work.
The range of $B{\left(E2; 4_1^+ \to 2_1^+\right)}$ values in \nuc{58}{Fe} agrees well with the current shell-model calculations, and is significantly lower than the previously adopted value derived from lifetime measurements~\cite{Nesaraja2010}.
These significantly lowered values reaffirm the effectiveness of the shell-model description of nuclear structure in the $N = Z = 28$ region.

Previously measured lifetimes in \nuc{58}{Fe} have been re-evaluated using updated stopping powers.
Lifetimes other than that of the $4_1^+$ state generally agree well with other previous measurements where they exist.
However, the re-evaluated $4_1^+$ lifetime is significantly longer than measured in previous works.
The $B{\left(E2; 4_1^+ \to 2_1^+\right)}$ value derived from this lifetime agrees well with both shell-model calculations and the present Coulomb-excitation measurement.
For DSAM measurements like that of Ref.~\cite{Bolotin1978}, where both electronic and nuclear stopping are important, deviations in stopping powers can result in small changes in the lifetimes of some states but large changes in other states.
It is clear that lifetimes that were evaluated based on LSS stopping powers should be carefully re-examined case-by-case to assess the impact of using updated stopping powers.

\begin{acknowledgments}
	The authors wish to acknowledge the excellent support of the technical staff of the Department of Nuclear Physics and Accelerator Applications.
	We thank J.~T.~H.~Dowie, P.~McGlynn, and Y.~Y.~Zhong for their contribution to data collection for this work.
	We also thank T.~Ratcliff for measuring the target thickness.
	J.~A.~Woodside, L.~J.~McKie, and N.~J.~Spinks acknowledge support of the Australian Government Research Training Program Scholarship.
	This material is based upon work supported by the Australian Research Council Grant No.\ DP210101201 and the International Technology Center Pacific (ITC-PAC) under Contract No.\ FA520919PA138.
	The authors acknowledge the facilities, and the scientific and technical assistance provided by Heavy Ion Accelerators (HIA).
	HIA is supported by the Australian Government through the National Collaborative Research Infrastructure Strategy (NCRIS) program.
	This material is based upon work supported in part by the U.S.\ Department of Energy, Office of Science, Office of Nuclear Physics under Contract No.\ DE-AC05-00OR22725.
	The publisher acknowledges the U.S.\ government license to provide public access under the DOE Public Access Plan (\url{http://energy.gov/downloads/doe-public-access-plan}).
\end{acknowledgments}

\bibliographystyle{apsrev4-2} 
\bibliography{fe58-prc}

\end{document}